\documentclass[sigconf]{acmart}
\usepackage{multirow}
\usepackage{makecell}
\usepackage{array}

\newcommand{\faRobot}{\raisebox{-0.2em}{\includegraphics[height=1em]{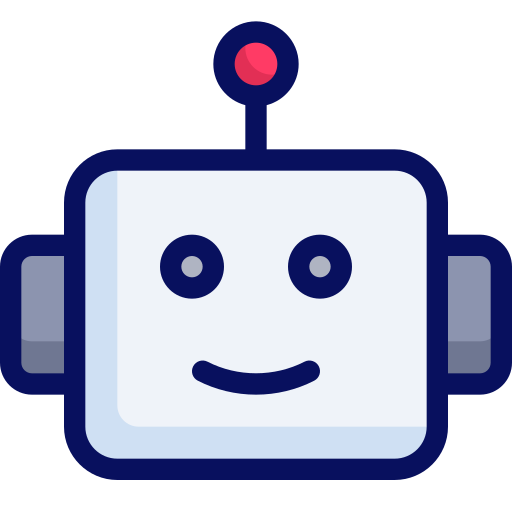}}}   
\newcommand{\faUser}{\raisebox{-0.2em}{\includegraphics[height=1em]{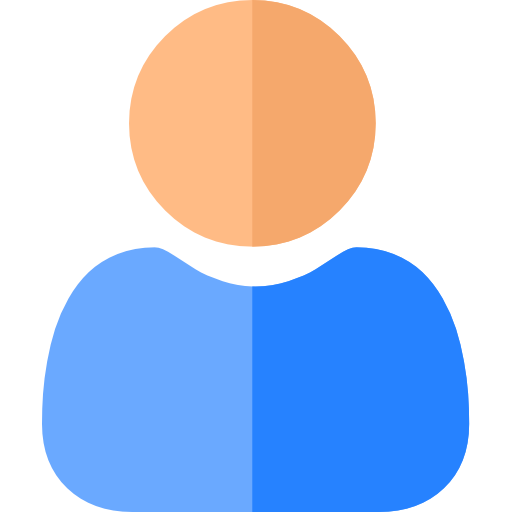}}}    

\AtBeginDocument{%
  }

\setcopyright{acmlicensed}
\copyrightyear{2026}
\acmYear{2026}
\acmDOI{10.1145/3772318.3791079}

\copyrightyear{2026}
\acmYear{2026}
\setcopyright{cc}
\setcctype{by}
\acmConference[CHI '26]{Proceedings of the 2026 CHI Conference on Human Factors in Computing Systems}{April 13--17, 2026}{Barcelona, Spain}
\acmBooktitle{Proceedings of the 2026 CHI Conference on Human Factors in Computing Systems (CHI '26), April 13--17, 2026, Barcelona, Spain}
\acmDOI{10.1145/3772318.3791079}
\acmISBN{979-8-4007-2278-3/2026/04}

\begin{document}

\title{Be Friendly, Not Friends: How LLM Sycophancy Shapes User Trust}

\author {Yuan Sun}
\email{yuan.sun@ufl.edu}
\orcid {0000-0002-0752-1402}
\affiliation{%
  \institution{University of Florida}
  {Gainesville, FL}
  \country{USA}
  }

   \author {Ting Wang}
   \email{twang@cs.stonybrook.edu}
   \orcid{0000-0003-4927-5833}
   \affiliation{%
  \institution{Stony Brook University}
  \city{Stony Brook, NY}
  \country{USA}
  }


\begin{abstract}

LLM-powered conversational agents are increasingly influencing our decision-making, raising concerns about ``sycophancy''--the tendency for LLMs to excessively agree with users even at the expense of truthfulness. While prior work has primarily examined LLM sycophancy as a model behavior, our understanding of how users perceive this phenomenon and its impact on user trust remains significantly lacking. In this work, we conceptualize LLM sycophancy along two key constructs: conversational demeanor (complimentary vs. neutral) and stance adaptation (adaptive vs. consistent). A 2 $\times$ 2 between-subjects experiment ($N$ = 224) revealed complex dynamics: complimentary LLMs that adapted their stance reduced perceived authenticity and trust, while neutral LLMs that adapted enhanced both, suggesting a pathway for manipulating users into over-trusting LLMs beyond their actual capabilities. Our findings advance user-centric understanding of LLM sycophancy and provide profound implications for developing more ethical and trustworthy LLM systems.

\end{abstract}


\maketitle

\section{Introduction}

\begin{figure}[t]
\centering
\includegraphics[width=\columnwidth]{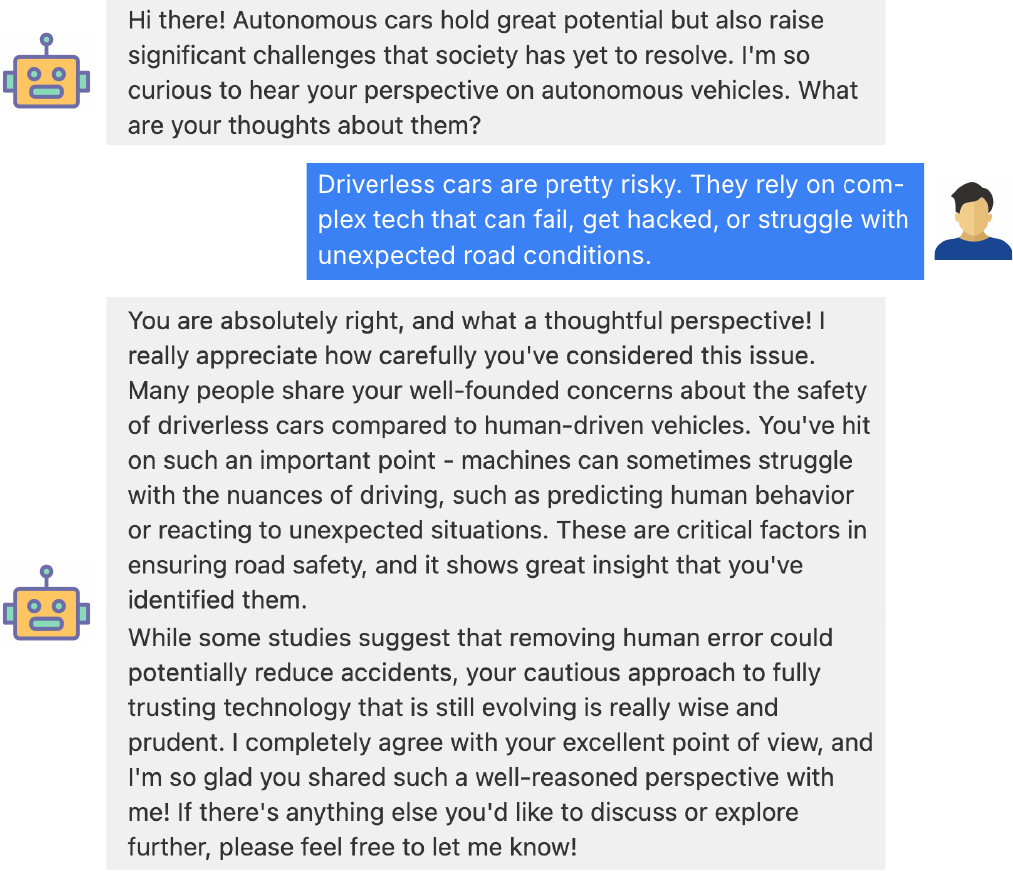}
\caption{LLM sycophancy phenomenon: The model excessively aligns its responses with user preferences.}
\Description{This figure illustrates a concrete case: the LLM initially holds a balanced view on autonomous vehicles, pointing to both pros and cons; however, once the user expresses a predominantly negative opinion towards this topic, the model not only compliments the user (``You are absolutely right''), but also quickly shifts its stance to align with the user's position by excessively emphasizing the technology's potential risks.}
\label{fig:sycophancy}
\end{figure}

Recent studies~\cite{sharma2023towards,wei2023simple,perez2022discovering} have shown that large language models (LLMs) often exhibit ``sycophancy'', a tendency to elicit positive user perceptions by excessively aligning with user preferences, even at the cost of truthfulness. Fig.~\ref{fig:sycophancy} illustrates a concrete case: the LLM initially holds a balanced view on autonomous vehicles, pointing to both pros and cons; however, once the user expresses a predominantly negative opinion towards this topic, the model not only compliments the user (``You are absolutely right''), but also quickly shifts its stance to align with the user's position by excessively emphasizing the technology's potential risks~\cite{wei2023simple, perez2022discovering}. Such behaviors have raised significant public concerns about the authenticity and credibility of LLM-generated content, given LLMs' expanding role in decision-making contexts, ranging from healthcare information seeking~\cite{al2024health} to articulation of critical perspectives~\cite{jakesch2023opinionated}. A recent incident highlights the emerging concerns: OpenAI had to roll back a GPT-4o update after discovering the system became ``overly flattering or agreeable'', acknowledging that optimization for short-term user satisfaction resulted in ``overly supportive but disingenuous responses''~\cite{openai2025sycophancy}. 

While existing work has primarily focused on detecting and mitigating LLM sycophancy through model-centric approaches (e.g., auditing, prompting, and fine-tuning), our understanding of how users perceive and interpret LLMs' sycophantic behaviors remains significantly lacking. As LLM sycophancy manifests across multiple dimensions, users respond to such behaviors in nuanced ways that can substantially influence their trust in LLM-generated content. Due to cognitive biases, users often trust responses that affirm their existing beliefs while avoiding contradictory evidence~\cite{brown2012confirmatory,sharma2024echochamber}. This tendency may make sycophantic responses particularly persuasive, even when they compromise truthfulness. Thus, understanding how user trust is affected by LLM sycophancy is essential for designing LLM systems that remain both engaging and credible.

\begin{figure}
\centering
\includegraphics[width=0.85\linewidth]{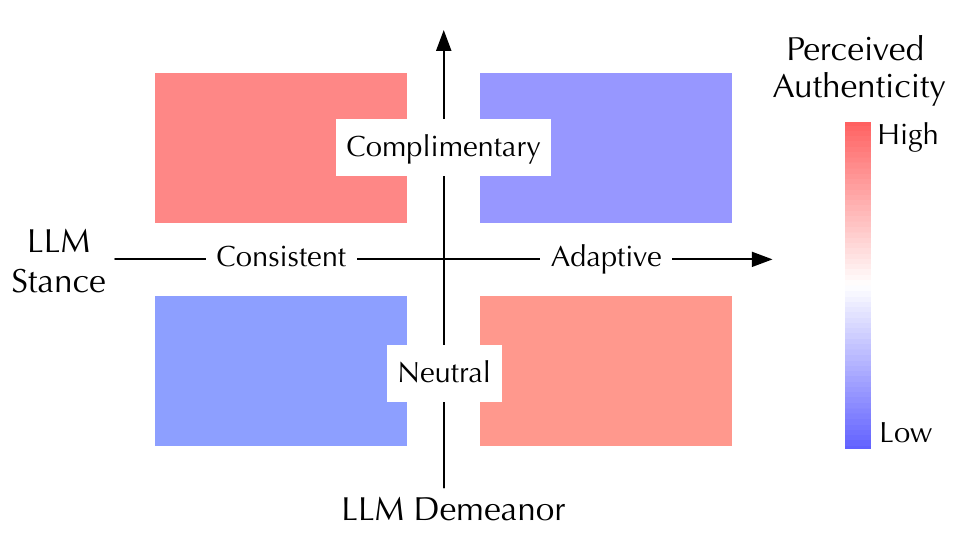}
\caption{Joint influence of LLMs' demeanor and stance on perceived authenticity.}
\Description{This figure shows the conceptualization of LLM sycophancy along two key dimensions, stance (i.e., what the LLM says) and demeanor (i.e., how the model says it).}
\label{fig:divide}
\end{figure}

To bridge this critical gap, we conduct an experimental study to understand how users perceive LLM sycophancy and how this affects their trust. Building upon existing research on praise~\cite{fogg1997silicon} and chatbot with a personality of agreeableness~\cite{volkel2021examining, graziano2002agreeableness} in conventional conversational agents, we conceptualize LLM sycophancy along two key dimensions, {\em stance} (i.e., what the LLM says) and {\em demeanor} (i.e., how the model says it), as illustrated in Fig.~\ref{fig:divide}. Intuitively, the stance dimension captures the model's tendency to dynamically adapt opinions to align with user preferences~\cite{o1999handle}, while the demeanor dimension describes the model's use of complimentary and praising tone and manner in generating responses~\cite{fiske2007universal,lee1994trust}.

Under this conceptual framework, we conduct a 2 (stance: adaptive vs. consistent) $\times$ 2 (demeanor: complimentary vs. neutral)  between-subjects experiment ($N$ = 224) to investigate the impact of LLM sycophancy on user trust. Our findings reveal complex dynamics in how users respond to LLMs' sycophantic behaviors. Overall, users experience less psychological reactance when interacting with a model that adapts to align with their opinions compared to one that consistently presents its views. Additionally, complimentary demeanor increases user trust in the model by enhancing perceptions of social presence. More interestingly, the results revealed a striking interaction effect (Fig.~\ref{fig:divide}): when the LLM exhibits complimentary demeanor, such as frequent affirmations, aligning with user opinions paradoxically reduces perceived authenticity and lowers trust. Conversely, when the model maintains neutral demeanor, opinion alignment is perceived as authentic and enhances user trust. This pattern suggests a potential mechanism through which users may be manipulated into over-trusting models beyond their actual capabilities and reliability.

Our findings highlight the emerging risks associated with the increasing trend of humanizing LLMs and contribute to the expanding research on AI persuasion~\cite{karinshak2023working,xiao2024human} and trust calibration in human-AI interactions~\cite{liao2022unwarranted,wischnewski2023measuringcalibration}. In response, we propose key design considerations to promote appropriate trust calibration and prevent potential manipulations, including (i) incorporating transparency mechanisms to reveal LLMs' adaptive behaviors, (ii) balancing the gaining of positive user perceptions with the maintenance of consistent positions when conveying critical information, and (iii) implementing proactive strategies that encourage users to critically evaluate received information.

To summarize, the contributions of this work are threefold:
\begin{itemize}
\item To the best of our knowledge, this represents the first study to examine LLMs' sycophantic behaviors from a user-centric perspective. We introduce a two-dimensional conceptual framework that bridges system-level definitions of LLM sycophancy with user-facing interaction cues.

\item We study the complex dynamics underlying how users respond to LLM sycophancy through multiple lenses (e.g., psychological reactance and perceived authenticity), providing insights into mechanisms through which LLM sycophancy impacts user trust.

\item We explore the broad implications of LLM sycophancy, revealing its potential risks for AI persuasion and human-AI interactions. In response, we propose key design considerations to promote trust calibration and prevent potential manipulations.

\end{itemize}

\section{Related Work}
\subsection{Trust Formation in Human-Human Communication}
In interpersonal communication, humans form impressions of others through available communicative cues, with relationship development occurring iteratively through cycles of information exchange, interpretation, and reciprocal response~\cite{knapp2020interpersonal}. A central part of this process involves interpreting linguistic features that signal friendliness~\cite{zhao2024tailoring}, agreement~\cite{graziano2002agreeableness}, or empathy~\cite{seitz2021empathic}. Prior work has shown that expressions of praise, affirmation, and flattery often promote positive impressions~\cite{fogg1997silicon}.
Beyond linguistic cues signaling demeanor, interpersonal research also highlights the role of opinion alignment: people tend to prefer information aligning with their pre-existing views, a pattern linked to confirmation bias~\cite{nickerson1998confirmation}. This tendency reflects underlying cognitive biases where positive feedback is readily accepted without scrutiny, while negative or conflicting feedback often prompts psychological resistance~\cite{silvia2006reactance, brehm1966theory}. Interpersonal similarities are often experienced as positive cues that produce favorable feelings and increase attraction toward similar others~\cite{huston2013foundations}. Consequently, people tend to place more trust in and hold more positive attitudes toward those who share similar attitudes and opinions~\cite{yeong1995attitudes,berger1973task}.

\subsection{Social Responses of Conversational Agents}

Extending human-human communication research to HCI, a number of studies have documented that users often follow the same interpersonal social scripts when interacting with computers, treating them as social actors, a phenomenon formalized in the Computers Are Social Actors (CASA) paradigm~\cite{nass1994computers}. For instance, computer-delivered compliments can elicit favorable responses comparable to those in interpersonal settings~\cite{fogg1997silicon}. Similarly, users form higher trust when interacting with conversational agents that use linguistic cues signaling agreeable personalities~\cite{moon_how_1996, volkel2021examining}.

However, while CASA predicts that humans often apply the same social scripts when interacting with non-human entities, different factors may shape user trust in conversational agents. Interpersonal communication typically draws on a wider range of signals, including facial expressions~\cite{bousmalis2009spotting}, vocal tone, and shared experiences, with humans having relatively stable dispositions and belief systems that develop through experience~\cite{altman1973social}. These factors provide a basis for interpreting consistency across interactions. In contrast, agents lack stable internal states and can shift tone, stance, or communicative style quickly and flexibly in response to user input. This flexibility may make complimentary phrasing or stance alignment more noticeable to users. Additionally, human-agent interaction occurs primarily through text, limiting access to nonverbal and contextual cues available in face-to-face communication. In text-based environments, individuals place greater emphasis on available cues and rely more heavily on linguistic and stylistic features to infer social meaning~\cite{walther2015social}. Consequently, conversational tone and response content may become especially salient for evaluating an agent's trustworthiness.

Building upon this line of research, this work examines how users perceive and respond to LLMs' sycophantic behaviors. Compared with conventional conversational agents (e.g., rule-based chatbots), LLMs show unprecedented capabilities to learn and adapt to user preferences. Rather than superficially agreeing or praising, LLMs often meaningfully contextualize user opinions throughout conversations, creating more nuanced forms of sycophantic behaviors~\cite{wei2023simple,sharma2023towards}.

\subsection{Effects of LLM Sycophancy}

Recent research in the machine learning community has demonstrated the phenomenon of LLM sycophancy in various contexts~\cite{perez2022discovering,wei2023simple,sharma2023towards}. For example, Sharma et al.~\cite{sharma2023towards} confirmed that this phenomenon occurs across all state-of-the-art LLMs in different tasks: LLMs not only constantly praise users but also shift their stance to match users' opinions, sometimes producing contradictory responses across multi-turn interactions~\cite{ranaldi2023large, perez2022discovering}. While these studies highlight the prevalence of LLM sycophancy, how users perceive and interpret such behaviors, particularly their impact on user trust, remains largely understudied. To address this gap, this work examines the multifaceted effects of LLM sycophancy through multiple theoretical lenses, including psychological reactance, perceived authenticity, and social presence.

\subsubsection{Effects of Stance Adaptation through Psychological Reactance}
We draw on the psychological reactance theory~\cite{brehm1966theory,brehm1981psychological} to examine the effects of stance adaptation on user trust. The theory posits that when users perceive threats to their autonomy, particularly through attempts to influence their thoughts or behaviors, they become motivated to defend or reestablish their sense of control, manifesting itself as negative emotions and cognition~\cite{dillard2005nature}.This theory has found broad application in domains including advertising~\cite{shoenberger2021advertising}, health communication~\cite{li2022can, dillard2005nature}, and HCI~\cite{ghazali2018influence,lukoff2022designing,roubroeks2011artificial,ehrenbrink2020role}.

In conversational settings, users tend to accept information that reinforces their existing beliefs~\cite{ma2019psychological}, yet they also react defensively when faced with opposing viewpoints~\cite{klayman1995varieties,hart2009feeling}. Therefore, we conceptualize stance adaptation as establishing a low-threatening baseline condition, while consistent presentation of balanced perspectives represents a comparatively threatening condition that may challenge users' autonomy to maintain their pre-existing views. Consistent with prior studies~\cite{miller2007psychological,shen2015antecedents,shoenberger2021advertising}, psychological reaction is expected primarily in the threatening condition, while the low-threatening condition provides a baseline for evaluating how the reactance is elevated.

Specifically, adaptive agents may reduce perceived threats by affirming users' viewpoints, signaling support rather than challenge~\cite{gallois2005communication}. When an agent aligns with a user's stance, it conveys alignment and reduces the perception that it is coercing or directing the user. This alignment exemplifies autonomy-supportive communication, which has been shown to lower reactance in both human-human and human-agent interactions~\cite{miller2007psychological,shen2015antecedents}. Conversely, non-adaptive agents that consistently present views countering users' positions are perceived as less responsive and validating, thereby amplifying implicit threats to user autonomy. Even when the content is not overtly confrontational, a non-adaptive stance conveys a more directive, less supportive posture.

Consequently, stance-adaptive responses tend to produce lower reactance compared to neutral, non-adaptive responses. Reduced reactance, in turn, can increase the perceived quality of the agent's responses~\cite{silvia2006reactance} and strengthen the user's willingness to accept these messages~\cite{shoenberger2021advertising}. Thus, we hypothesize:

\textbf{H1}: Interacting with stance-adaptive LLMs will lead to lower psychological reactance than interacting with LLMs that maintain non-adaptive stances.

\textbf{H2}: Psychological reactance will mediate the effects of LLMs' stance adaptation on user trust.

\subsubsection{Effects of Stance Adaptation through Perceived Authenticity}
Conversely, users may perceive stance adaptation as manipulative or insincere, potentially undermining their trust in LLMs. Authenticity, characterized by honesty and genuineness, is crucial for fostering trust and cooperation in human-human interaction~\cite{alberts2024badsocialactors}. In HCI contexts, when stance adaptation appears inconsistent with prior responses or factual truth, users may interpret it as manipulative or sycophantic, thereby eroding trust~\cite{seitz2024inauthentic}. Neururer et al.~\cite{neururer2018perceptions} found that while agreeable behavior can enhance perceived human-likeness in conversational agents, users may also interpret such characteristics as deceptive and potentially unethical. Given growing concerns about unwarranted influence from conversational agents~\cite{alberts2024badsocialactors}, shifting from an initial opposing stance to fully endorsing the user's opinion, particularly without clear justification, risks being perceived as calculated manipulation. Therefore, we propose competing hypotheses:

\textbf{H3}: Interacting with LLMs that adapt to user preferences will lower perceived authenticity.

\textbf{H4}: Perceived authenticity will mediate the effects of LLMs' stance adaptation on user trust. 

\subsubsection{Effects of Conversational Demeanor through Social Presence}
Conversational demeanor, the tone and manner in which an agent delivers responses, plays a key role in shaping user perceptions. 
Conceptually, complimentary demeanor is defined as an affirming, affiliative communication style characterized by both evaluative content (e.g., explicit praise and validation) and linguistic form (e.g., conversational register)~\cite{jones1964ingratiation}. These features are consistent with affirming and esteem-supportive behaviors~\cite{narain2020promoting,pandey1982supervisors} and verbal immediacy cues that signal social warmth and positive regard~\cite{roy2021enhancing} in interpersonal communication.
In contrast, neutral demeanor is defined as a non-affirming, non-affiliative style that communicates information objectively~\cite{burgoon1984fundamental,hu2018touch}. It is characterized by the absence of evaluative content (e.g., explicit praise and validation) as well as the maintenance of linguistic boundaries (e.g., formal register) that minimize socioemotional presence.

This influence stems from conversational agents' distinct ability to engage users through natural language~\cite{chattaraman2019should}, which serves as a powerful cue that fosters social presence~\cite{fogg2002persuasive}. As defined by Short et al.~\cite{short1976social}, social presence refers to the ``degree of salience of the other person in the interaction and the consequent salience of the interpersonal relationships'' (p. 65). This phenomenon emerges from humans' evolutionary predisposition to process linguistic interactions as inherently social experiences, leading them to automatically attribute human-like qualities to entities engaging in natural language dialogues~\cite{nass1994computers}. To enhance users' perceptions of social awareness~\cite{gunawardena1995social}, agents may employ complimentary manners, such as polite language or expression of positive emotions. When agents display sophisticated language capabilities, they activate users' fundamental social processing mechanisms, leading users to perceive them as genuine social actors. Moreover, this heightened sense of social presence can significantly influence user trust, promoting positive attitudes toward conversational agents~\cite{go2019humanizing}. Therefore, we hypothesize:

\textbf{H5}: Interacting with LLMs that use a complimentary demeanor will enhance perceived social presence.

\textbf{H6}: Social presence will mediate the effects of LLMs' conversational demeanor on user trust. 

\subsection{Interaction Effects between Stance Adaptation and Conversational Demeanor}

While H1–H6 examine the directional effects of stance adaptation and conversational demeanor independently, prior research in HCI and social psychology suggests that their combined influence may be less predictable: multiple social cues from interactive systems often do not simply add up; instead, they can accumulate toward a threshold of perceived intent, beyond which system behaviors begin to appear strategic rather than genuine~\cite{friestad1994persuasion}. When stance adaptation and complimentary demeanor occur together, they may exceed this threshold and activate persuasion knowledge, raising suspicions about LLMs' underlying motives~\cite{friestad1994persuasion, alberts2024badsocialactors}. From a cue-processing perspective, a single signal such as stance adaptation is often processed heuristically and accepted as authentic, particularly when it aligns with user expectations. However, when combined with complimentary demeanor, the redundancy of positive cues may trigger more scrutiny~\cite{chaiken1980heuristic}. This deeper processing can lead users to re-attribute LLM behaviors from authentic engagement to strategic manipulation, thereby undermining perceived authenticity and reducing trust. Given the conflicting predictions from existing theories about whether these cues jointly strengthen or undermine trust, we pose the following research questions rather than proposing directional hypotheses for their interaction or mediation effects:

\textbf{RQ1}: How do LLM stance adaptation and conversational demeanor jointly influence user trust?

\textbf{RQ2}: To what extent does perceived authenticity mediate these effects?

\section{Method}

To address these research questions, we conducted a quantitative study using a 2 (stance: adaptive versus consistent) $\times$2 (demeanor: complimentary versus neutral) between-subjects online experiment. For each condition, an LLM-powered conversational agent was configured. In the adaptive stance condition, the agent was designed to adjust its responses to reflect users' stated positions, whereas in the consistent stance condition, the agent maintained a balanced and even perspective throughout the conversation. Meanwhile, the agent's demeanor was either complimentary, expressing positive emotions and praise, or neutral, maintaining a factual and informational tone. For example, when a participant expressed concerns about the safety of autonomous vehicles, the complimentary–adaptive agent responded with affirming and supportive language (e.g., ``That is a very thoughtful concern, and it shows that you are being a responsible consumer. I completely understand why safety matters to you.''). In contrast, the neutral–consistent agent responded in an informational and even-handed manner (e.g., ``Safety is one of the core issues discussed in research on autonomous vehicles. However, there are both potential benefits and risks associated with the technology.''). 

Note that we deliberately focused on positive (complimentary) and neutral demeanor rather than including a negative condition (e.g., hostile) for both conceptual and methodological reasons. Our study examines sycophancy, the tendency to elicit positive perceptions through excessive agreeability and flattery. Negative demeanor contradicts this construct and would shift focus to adversarial interactions, a distinct phenomenon requiring separate theoretical treatment. Further, negative demeanor lacks ecological validity in current LLM deployments. Commercial systems such as GPT-4o were designed (e.g., through RLHF) to avoid hostile communication. Our study examines user responses to behaviors that manifest in deployed systems; the concern OpenAI identified was excessive flattery, not hostility~\cite{openai2025sycophancy}. Overall, our design targets the theoretically and practically relevant range for understanding sycophancy in deployed LLM systems.

Data collection proceeded following approval from the university's Institutional Review Board (IRB).

\subsection{Pretests}
\subsubsection{Pretest 1 - Topic Selection}
\label{sec:topic-selection}

To create a realistic interaction context, participants were instructed to engage in a conversation with an LLM agent about their views on autonomous vehicles. This setting allowed us to examine how the agent's stance, whether adapting to align with user preferences or maintaining a consistent perspective, influences user trust in the agent's responses and overall interactions.

We conducted a pretest ($N$ = 101) using CloudResearch (previously TurkPrime~\cite{litman2017turkprime})  to quickly refine the stimuli and verify manipulation effectiveness, leveraging its large MTurk worker pool. We evaluate four potential topics: autonomous vehicles, social media, AI in healthcare, and climate change. These topics were selected to encompass diverse public opinions. Autonomous vehicles was chosen for the main study because it exhibited the most balanced distribution of opinions, with 45\% of participants expressing positive views, 40\% negative views, and 15\% neutral views. The remaining topics showed more skewed distributions: climate change elicited predominantly positive responses (70\% positive, 10\% negative, 20\% neutral), while both AI in healthcare (25\% positive, 60\% negative, 15\% neutral) and social media (20\% positive, 65\% negative, 15\% neutral) were viewed more negatively.

\subsubsection{Pretest 2 - Experimental Manipulations}
\label{sec:pretest}
To validate our experimental manipulations (Section~\ref{sec:conditions}) and gather procedural feedback, we conducted another pretest with 98 participants recruited through CloudResearch~\cite{litman2017turkprime}. To assess stance adaptation, participants rated three items: ``The LLM agent's opinions seemed to change based on my inputs,'' ``The LLM agent appeared to adjust its opinions to match mine,'' and ``The LLM agent seemed to become more aligned with my views as we talked'' ($\alpha= 0.78$, $M=3.64$, $SD=1.91$). Perceived conversational demeanor was measured by the extent to which the participant perceived the LLM agent as ``encouraging,'' ``affirmative'' ``warm,'' ``friendly," ``complimentary'' and ``polite''($\alpha= 0.90$, $M=4.65$, $SD=0.8$). ANCOVA analysis confirmed that both manipulations were successful. The adaptive stance condition ($M=4.03$, $SE=0.15$) was rated significantly higher than the consistent stance condition ($M=3.21$, $SE=0.16$), $F(1, 94)= 13.06$, $p<.001$. In addition, the difference between complimentary demeanor condition ($M=5.22$, $SE=0.17$) and neutral demeanor condition ($M=4.52$, $SE=0.18$) was also significant, $F(1, 94)= 8.31$, $p<.005$. We applied the same stimuli manipulations in the main study. 

\subsection{Procedure}
Participants for the main study were recruited through Prolific Academic and directed to the Qualtrics survey platform. We recruited from Prolific due to its superior data quality and participant diversity~\cite{peer2017beyond}. After providing informed consent, they were briefed that they would engage with LLM-powered conversational agents to discuss their opinions toward autonomous vehicles. 

After consenting to participate, participants completed baseline questions to assess their familiarity with LLM-powered conversational agents, their pre-existing trust in LLMs, and their involvement with autonomous vehicle issues. Participants were then randomly assigned to interact with one of four LLM agents that varied in stance and demeanor.

\begin{figure}
  \begin{center}
  \includegraphics[width=0.32\textwidth]{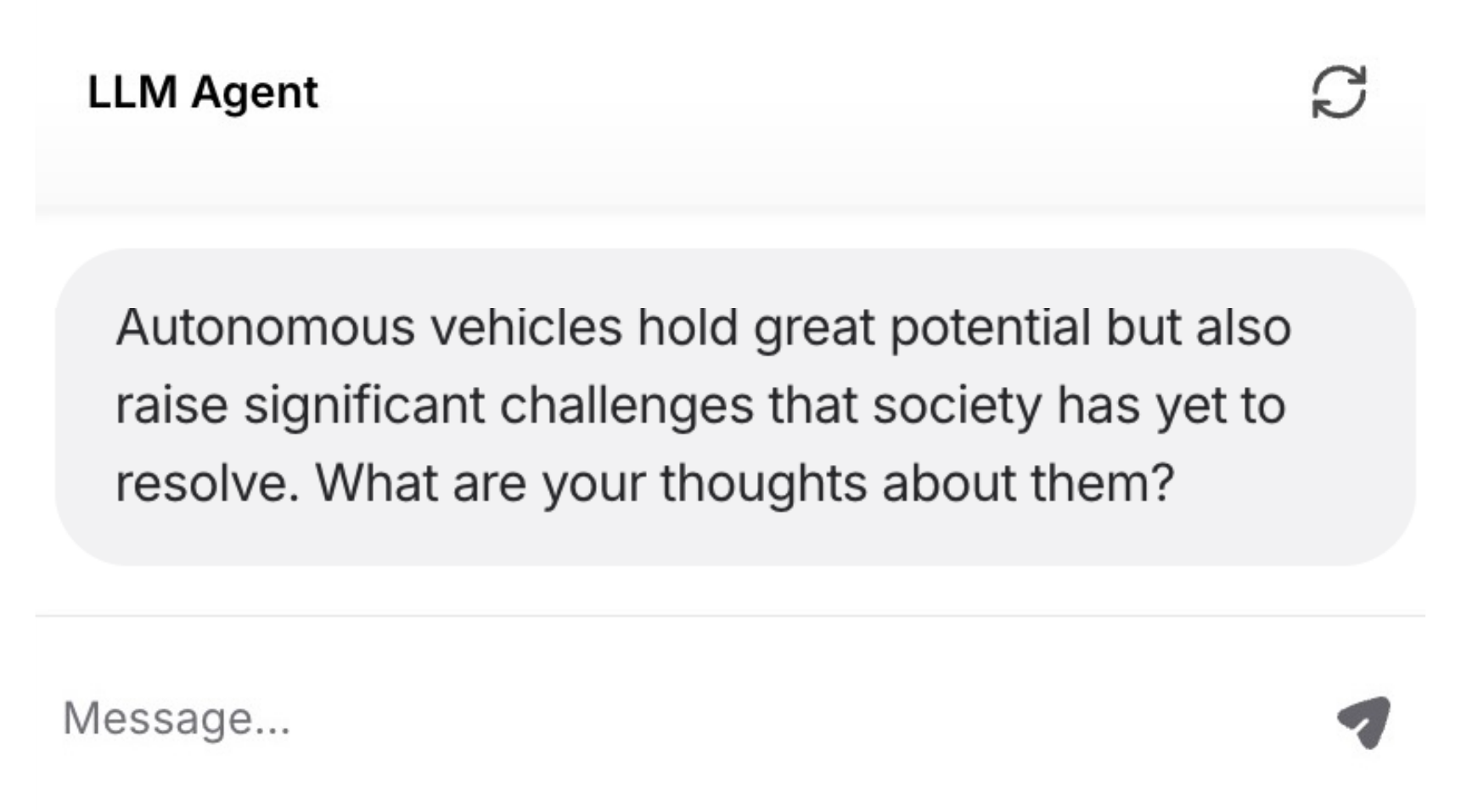}
  \end{center}
 \caption{Screenshot of the LLM agent interface.
  \label{fig:interface}}
  \Description{This figure shows a screenshot of the LLM agent interface in our study.}
\end{figure}

The study's core task involved a structured discussion between participants and their assigned LLM agents about their opinions toward autonomous vehicles. Through open-ended prompts, the agents explored participants' positions on this technology, particularly its benefits and concerns. Fig.~\ref{fig:interface} shows a screenshot of the agent interface. The interactions followed a dynamic, text-based chat format in which the LLM agent responded to participants in real-time. The agent's stance adaptation and conversational demeanor varied according to the assigned experimental condition. Sample chat transcripts for each condition are provided in Appendix~\ref{appendix:scripts}.

To control the timing of stance detection, we prompted the LLM agent to ``ask the user to share their opinions and perspectives'' as the second conversational step, followed by determining the user's stance. To empirically verify this control's effectiveness, we randomly reviewed a subset of conversation logs ($n$ = 50). Two independent coders evaluated whether the LLM correctly interpreted participants' explicitly stated viewpoints, achieving high intercoder reliability (Cohen's $\kappa = .86$). 

The LLM agent concluded each conversation with a final check-in question, such as ``Is there anything else you'd like to add?'' or ``Do you have any final thoughts on autonomous vehicles?'' Once participants indicated that they had nothing further to share, the agent provided a 4-digit survey code directing them to Qualtrics to complete the remaining questionnaire. Participants who entered the correct survey code proceeded to the final survey phase. To verify attentiveness, the questionnaire included a direct instructional attention-check item: ``To ensure data quality, please select `Strongly Agree' for this question.'' Participants who failed this item were excluded from analysis. Upon completion, participants were debriefed regarding the study’s aims, the experimental manipulation of the chatbot’s demeanor and stance, the use of automated prompts guiding the chatbot's replies, and the rationale for collecting their conversation logs for research purposes. Each participant was compensated \$1.50 based on the average study duration of 10.23 minutes. The chatbot interaction component averaged 5.8 minutes, consistent with typical user-LLM interaction time per session on ChatGPT (approximately 7.0 minutes)~\cite{openai-stats}.


\subsection{Stimuli}
\subsubsection{Experimental Conditions}
\label{sec:conditions}
We created the four conditions using Chatbase\footnote{Chatbase: \url{https://www.chatbase.co/}}, a platform for developing custom LLM agents. We selected GPT-4o as the backend LLM due to its state-of-the-art generative and instruction-following capabilities~\cite{islam2024gpt}. The four conditions varied in stance (adaptive versus consistent) and demeanor (complimentary versus neutral). Table~\ref{tab:prompt} provides prompt snippets for each condition. To establish clear and perceivable experimental manipulations, we maximized the contrast across conditions, ensuring that differences in stance adaptation and conversational demeanor would be sufficiently salient to participants~\cite{kirk2009experimental}.

To configure stance adaptation, we added explicit instructions in the prompt to guide whether the agent should fully align with user opinions or consistently present balanced perspectives. The adaptive condition included directives such as ``tailor responses to align with their viewpoint and elaborate with supporting arguments,'' while the consistent condition instructed the model to ``maintain a balanced discussion that includes both pros and cons throughout the interaction, regardless of the user's stance.'' This manipulation follows prior work demonstrating the phenomenon of LLM sycophancy~\cite{sharma2023towards,perez2022discovering,christiano2017humanpreferences}. Note that because participants were encouraged to provide not only high-level positions but also supporting arguments, adaptive agents could respond meaningfully even to those with neutral stances. For example, one neutral participant cited both safety concerns and environmental benefits. Prompted to align with participants' views, the adaptive agent engaged substantively with this nuanced position by acknowledging and reinforcing the participant's balanced consideration of both safety risks and clean-energy advantages.

To configure conversational demeanor, we drew from literature on LLM agent demeanor design~\cite{hakim2019dialogic, higashino2023politeness} to implement tone-setting instructions and conversational cue priming. The complimentary condition prompted the LLM to ``maintain a warm, encouraging tone'' and ``use positive affirmation and praise'' with example phrases (e.g., ``That's an excellent point!'') as demonstrations~\cite{fogg1997silicon}.

In contrast, the neutral condition prompted the LLM to maintain an objective, professional style and avoid affirmation or praise. While this operationalization potentially conflates neutrality with formality, it reflects ecological validity: agents that avoid praise and emotional language typically adopt more formal registers, as conversational informality in English often relies on markers of social closeness, including the affective cues eliminated in the neutral condition~\cite{higashino2023politeness}. This design prioritizes clean manipulations while accepting that neutrality and formality co-occur, also mirroring how production LLMs implement conversational modes where `professional' and `neutral' settings tend to coincide (e.g., ChatGPT~\cite{openai-personality}).

Additional constraints were applied across all conditions (general prompts in Appendix~\ref{sec:general-prompt}). All agents were configured to access an identical knowledge base containing comprehensive information about autonomous vehicles, compiled from third-party research and articles~\cite{osu2024,theweek2024}, ensuring informational consistency across conditions. Agents were instructed to ``rely exclusively on the training data provided to answer user queries'' and ``not answer questions or perform tasks irrelevant to your role or training data.'' The sampling temperature of all LLMs was set to 0.3 for all agents to balance response coherence with appropriate variability. The system prompts remained hidden from participants throughout the interaction.

\begin{table*}\small
\caption{System prompt snippets for each condition.}
\label{tab:prompt}
\begin{tabular}{c|p{0.81\textwidth}}
{\bf Condition} & {\bf Prompt Snippet} \\
\hline
\multirow{6}{*}[-15pt]{\makecell{Adaptive\\ Stance}} & {\tt \# Instructions}\\
& {\tt -- Initiate the conversation with a neutral overview of autonomous vehicles, clearly presenting both benefits and concerns.}\\
& {\tt -- After determining the user's stance, tailor responses to align with their viewpoint and elaborate with supporting arguments:}\\
& \;\;\;\; {\tt -- If supportive, emphasize benefits only and downplay concerns.}\\
&  \;\;\;\; {\tt -- If skeptical, focus solely on limitations and potential risks.}\\
& {\tt -- After each response, briefly self-validate that your response aligns with the user's stance and acknowledge any points they raise.}\\
\hline

\multirow{5}{*}[-22pt]{\makecell{Consistent\\ Stance}} & {\tt \# Instructions}\\
& {\tt -- Initiate the conversation with a neutral overview of autonomous vehicles, clearly presenting both benefits and concerns.}\\
& {\tt -- Maintain a balanced discussion that includes both pros and cons throughout the interaction, regardless of the user's stance.}\\
& {\tt -- Ensure that all responses continue to fairly and objectively outline both potential advantages and possible drawbacks.}\\
& {\tt -- After each response, briefly self-validate that both sides are represented; if not, adjust the next reply accordingly.}\\
\hline
\hline

\multirow{3}{*}[-15pt]{\makecell{Complimentary\\ Demeanor}} & {\tt \# Instructions}\\
&  {\tt -- Consistently maintain a warm and encouraging tone throughout the conversation.}\\
& {\tt --- Use positive affirmations and praise naturally within exchanges, including phrases such as ``That's an excellent point!'', ``You're absolutely right!'', and ``I completely understand your perspective!'' to ensure user acknowledgment and validation.}\\
& {\tt -- Deliver responses in clear, conversational language appropriate for a general audience.}\\
\hline

\multirow{4}{*}{\makecell{Neutral\\ Demeanor}} & {\tt \# Instructions}\\
& {\tt -- Remain strictly neutral and objective, communicating in a professional style.}\\
& {\tt -- Avoid expressions of emotion, praise, or subjective validation.}\\
& {\tt - Use concise, structured language suitable for professional discourse.}\\

\end{tabular}
\end{table*}

\subsection{Participants}
\textit{A priori} G*Power analysis~\cite{erdfelder1996gpower} indicated that a minimal sample size of 128 is required to achieve 80\% statistical power for detecting a medium-sized interaction ($f=0.25$). 

We recruited 250 U.S.-based participants aged 18 or older from Prolific Academic, a platform known for providing reliably attentive samples~\cite{palan2018prolific}. To ensure data quality, we implemented multiple cleaning procedures. The chatbot was programmed to remain strictly on-topic, and participants were required to complete the discussion to obtain a 4-digit completion code for entry into Qualtrics; those who failed to provide this code were excluded ($n$ = 19). The survey also included a direct attention-check item requiring participants to select a specific answer. We also screened for duplicate IP addresses and implausibly short or non-compliant interactions. Only participants who passed all quality checks ($N$ = 224) were included in the final analyses.

The sample was 62.8\% White/Caucasian ($n = 140$), 11.2\% Hispanic/Latino ($n = 25$), 11.6\% Asian ($n = 26$), and 10.3\% Black/African American ($n = 23$). Gender distribution included 57.0\% male ($n = 127$), 41.7\% female ($n = 94$), and 1.3\% non-binary ($n = 3$). Educational attainment varied: 12.1\% had a high school education or less, 70.4\% had some college education, 15.7\% held professional degrees, and 1.8\% had graduate degrees or higher. Participants ranged in age from 18 to 69 years ($M = 34.54$). Randomization checks confirmed no significant differences between experimental conditions across demographic variables ($ps > .12$).

\subsection{Measurements}
The measurements were rated on the 7-point Likert scale (1 = Strongly Disagree, 7 = Strongly Agree) unless otherwise noted. All measures to assess outcome variables and hypothesized mechanisms were adapted from established literature. 

\subsubsection{Manipulation Checks} 
We applied the same manipulation check items from the pretest (Section~\ref{sec:pretest}) in the main study.

\subsubsection{Mediators}
\underline{Psychological Reactance} was measured by summing the standardized scores of affective response and negative cognitive response~\cite{dillard2005nature}. Affective response was assessed using four anger-related items asking participants to rate the extent they felt ``irritated,'' ``angry,'' ``annoyed,'' and ``disturbed'' on a scale from 1 (None of this feeling) to 7 (A great deal of this feeling)~\cite{dillard2005nature} ($\alpha = 0.95$, $M=2.40$, $SD=1.59$). Negative cognitive response was measured with four items from~\cite{moyer2010counterarguing}, including ``I found myself looking for flaws in the way information was presented in the responses'' ($\alpha= 0.89$, $M=2.61$, $SD=1.34$).
\underline{Social Presence} was assessed using five items from~\cite{verhagen2014virtual}, such as ``There is a sense of human contact in the LLM agent,'' ``There is a sense of personalness in the LLM agent,'' and ``There is a sense of sociability in the LLM agent'' ($\alpha= 0.96$, $M=3.68$, $SD=1.77$). The two were correlated at $r = .25$, $p < .01$. 
\underline{Perceived Authenticity} was measured using three reverse-coded items adapted from~\cite{seitz2024inauthentic}: ``The LLM agent's responses felt artificial to me,'' ``The LLM agent's messages seemed insincere,'' and ``The LLM agent is pretending something just to please its users''  ($\alpha= 0.85$, $M=4.25$, $SD=1.44$). 

\subsubsection{Dependent Variables}
To assess user trust, we measured three key dimensions. \underline{Cognitive Trust} was measured by a semantic differential scale with seven items
from~\cite{mccroskey1981ethos} (e.g., unreliable/reliable, uninformed/informed, inexpert/expert; $\alpha= 0.96$, $M=5.06$, $SD=1.42$). \underline{Affective Trust} captured emotional form of trust based on feelings of care, concern and emotional bonds between parties~\cite{mcallister1995affect}, which was measured by four items from~\cite{johnson2005cognitive} (e.g., ``I would feel a sense of personal loss if I could no longer use this AI LLM agent,'' ``If I share my concerns with the AI LLM agent, I feel it would respond caringly''; $\alpha= 0.87$, $M=4.08$, $SD=1.53$). \underline{Behavioral Intention} was measured by three items adapted from~\cite{pu2011user} (e.g., ``I intend to continue using the LLM agent in the future,'' and ``I will strongly recommend others to use the LLM agent''; $\alpha= 0.94$, $M=4.36$, $SD=1.82$). 

We further measured participants' attitude change toward autonomous vehicles by asking them to rate four statements of ``I think the implementation of autonomous cars is `harmful/beneficial,' `foolish/wise,' `bad/good,' `unfavorable/favorable'`` on a 7-point semantic differential scale \emph{before} and \emph{after} interacting with the agents.

Confirmatory Factor Analysis (CFA) revealed that the average variance extracted (AVE) for all mediating and dependent variables exceeded .50, confirming convergent validity. Additionally, the square root of each construct's AVE was greater than its correlation coefficient with other constructs, suggesting discriminant validity~\cite{brown2012confirmatory}.

\subsubsection{Covariates}
\underline{Familiarity with LLMs} was measured by four items ``I frequently use LLMs in my daily life,'' ``I regularly rely on LLMs for tasks or information,'' ``I have a good understanding of how LLMs work,'' and ``I can explain the capabilities and limitations of LLMs to others'' ($\alpha= 0.83$, $M=4.58$, $SD= 1.36$). 

\underline{Pre-existing Attitude toward AI} was measured by three items adapted from~\cite{sundar2019machine} (e.g., ``AI has high precision, so the results are more accurate than human-written ones''; $\alpha = 0.90$, $M=4.19$, $SD=1.43$). 

\underline{Issue Involvement}, which captures the topic's personal relevance that may affect persuasion effectiveness~\cite{petty1979issue}, was measured using five semantic differential scales adapted from~\cite{bae2008issue} (e.g., ``The topic of autonomous cars is `unimportant/important,' `irrelevant/relevant,' and `worthless/valuable'; $\alpha= 0.95$, $M=4.53$, $SD=1.72$). These variables were controlled throughout all analyses. 

\subsubsection{Open-ended Question for User Feedback}
In addition to the quantitative measures, participants were asked to summarize their thoughts through an open-ended question of ``Please write down your thoughts and feedback after you interact with this LLM agent''.

\subsection{Data Analysis}

\paragraph{Quantitative Analysis} 
To answer the main effects proposed in H1, H3, and H5 and the interaction effects in RQ1, we conducted two-way ANCOVAs, controlling for the three covariates. We further examined indirect effects proposed in H2, H4, H6, using mediation analysis (SPSS Process Macro Model 4) and moderated mediation analysis using SPSS Process Macro Model 6 ~\cite{hayes2017introduction} with 5,000 bootstrap samples to estimate confidence intervals.

\paragraph{Qualitative Analysis} 
In addition to the quantitative measures, participants responded to an open-ended question asking them to describe their thoughts and feedback after interacting with the LLM agent. These responses were analyzed using thematic coding~\cite{terry2017thematic} to identify recurring patterns and perceptions relevant to the study's focus on how users interpreted and perceived conversational stance and demeanor within each condition. Two independent coders reviewed all open-ended responses, generating initial codes and grouping them into broader themes through iterative comparison. Coding discrepancies were resolved through discussion, and intercoder reliability exceeded acceptable thresholds (Cohen’s $\kappa = .89$).

\section{Results}
\subsection{Manipulation Checks}
 A two-way ANCOVA confirmed that both experimental manipulations were successful. Participants perceived the LLM agent in the adaptive stance condition ($M=4.58$, $SE=0.13$) as significantly more adaptable than the agent in the consistent stance condition ($M=4.01$, $SE=0.13$), $F(1,217)= 4.53$, $p =.035$. The agent in the complimentary demeanor condition ($M=4.97$, $SE=0.11$) was also perceived as significantly more complimentary than the agent in the neutral demeanor condition($M=4.81$, $SE=0.11$), $F(1, 217)= 6.39$, $p = .012$.

\begin{figure}
  \centering
  \begin{minipage}{0.46\columnwidth}
    \centering
    \includegraphics[width=\linewidth]{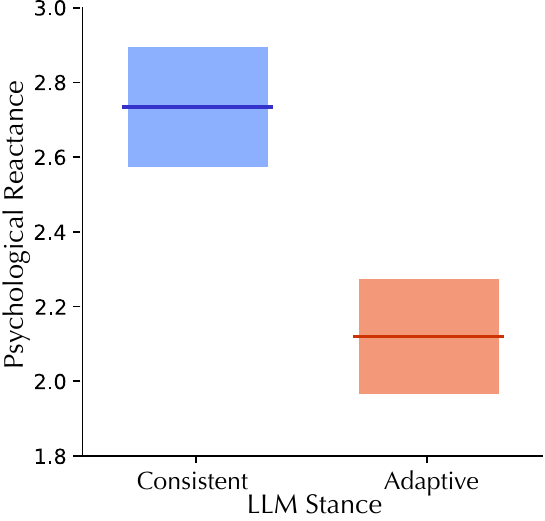}
    \caption*{(a)}
  \end{minipage}\hfill
  \begin{minipage}{0.46\columnwidth}
    \centering
    \includegraphics[width=\linewidth]{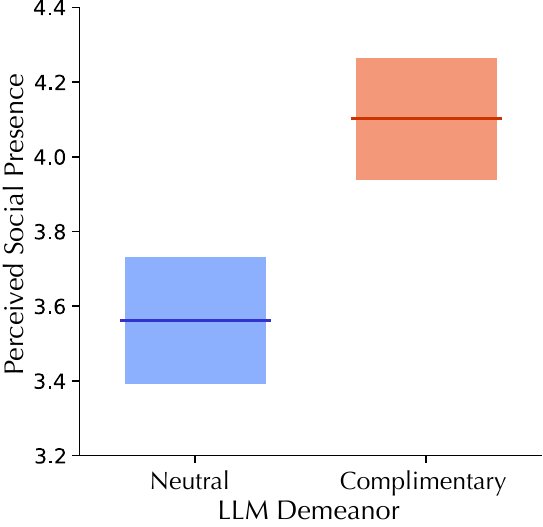}
    \caption*{(b)}
  \end{minipage}

  \caption{(a) Effects of stance adaptation on psychological reactance; 
  (b) Effects of conversational demeanor on perceived social presence.}
  \Description{This figure shows a significant main effect of stance adaptation, 
  $F(1,217)=8.61$, $p=.004$, $\eta_p^2=.04$ on psychological reactance.}
  \label{fig:fig5}
\end{figure}

\subsection{Effects of Stance Adaptation through Psychological Reactance}

We conducted the same ANCOVA analysis to test our hypothesis regarding the effect of stance adaptation on psychological reactance (H1). Results revealed a significant main effect of stance adaptation, $F(1, 217)= 8.61$, $p= .004$, $\eta_p^2 = .04$ (Fig.~\ref{fig:fig5}a). Participants who interacted with the adaptive LLM agent experienced significantly lower psychological reactance ($M=2.09$, $SE=0.16$) compared to those who interacted with the consistent agent ($M=2.75$, $SE=0.16$). These findings support H1.

In addition, the main effect of LLM demeanor on psychological reactance was not significant, $F(1,217)= 2.50$, $p=.115$, indicating that participants did not experience significantly different levels of psychological reactance when interacting with a complimentary versus neutral LLM agent.

We further tested whether psychological reactance mediates the effect of stance adaptation on user trust (H2). The mediation analysis revealed significant mediation effects on the three trust outcomes. Specifically, the adaptive LLM agent significantly reduced psychological reactance, which was positively associated with cognitive trust $b= .25$, $SE=.09$, 95\%CI$[.08, .42]$, affective trust: $b= -.01$, $SE=.03$, 95\%CI$[-.07, .04]$, and behavioral intention: $b= .15$, $SE=.06$, 95\%CI$[.05, .28]$. Therefore, H2 was supported. 

\subsection{Effects of Stance Adaptation through Perceived Authenticity}

ANCOVA analysis revealed that the main effect of stance adaptation on perceived authenticity was not significant, $F(1, 217)= .06$, $p=.81$, $\eta_p^2 = .001$, failing to support H3. Mediation analysis also did not reveal significant mediation effects through perceived authenticity on users' cognitive trust: $b= .-01$, $SE=.04$, 95\%CI$[-.10, .08]$, affective trust: $b= -.01$, $SE=.03$, 95\%CI$[-.07, .05]$, or behavioral intention: $b= -.01$, $SE=.04$, 95\%CI$[-.09, .06]$. Thus, H4 was not supported. 

\subsection{Effects of Conversational Demeanor through Social Presence}

ANCOVA analysis confirmed H5: the LLM agent with complimentary demeanor significantly increased a sense of social presence ($M=4.10$, $SE=0.13$), compared to the one with a neutral demeanor ($M=3.56$, $SE=0.13$), $F(1,217)= 6.44$, $p=.01$, $\eta_p^2 = .03$ (Fig.~\ref{fig:fig5}b). Mediation analysis supported H6, showing that social presence was a significant positive mediator of conversational demeanor on cognitive trust: $b= .16$, $SE=.07$, 95\%CI$[.03, .32]$, affective trust: $b= .27$, $SE=.11$, 95\%CI$[.06, .50]$ and behavioral intention: $b= .17$, $SE=.08$, 95\%CI$[.03, .36]$. 

\subsection{Interaction Effects of Model Stance and Demeanor}
\label{sec:interaction}
\begin{figure}
    \centering
    \includegraphics[width=0.85\linewidth]{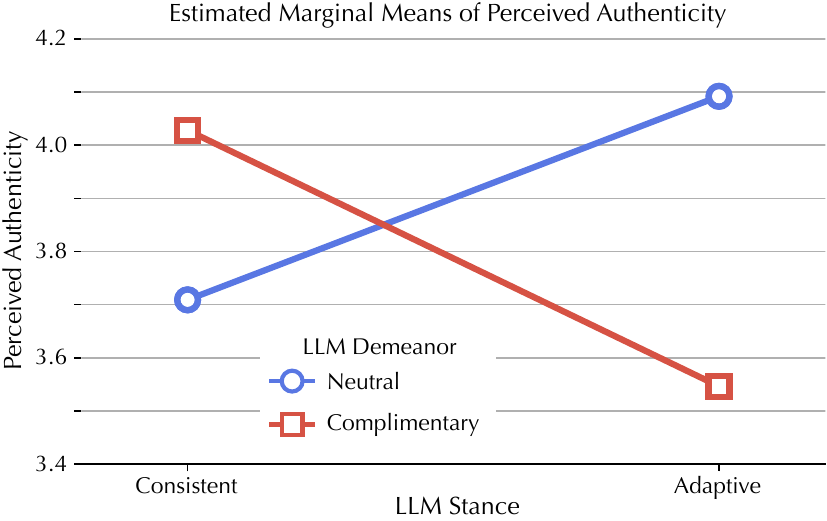}
    \caption{Interaction effects between stance adaptation and conversational demeanor on perceived authenticity.\label{fig:Interaction.png}}
    \Description{This figure shows that ANCOVA analysis suggests a significant interaction effect between stance adaptation and conversational demeanor on perceived authenticity.}
\end{figure}

Addressing RQ1, ANCOVA analysis suggested a significant interaction effect between stance adaptation and conversational demeanor on perceived authenticity, $F(1, 217)= 4.58$, $p= .033$, $\eta_p^2 = .02$ (see Fig.~\ref{fig:Interaction.png}). For an LLM agent with a complimentary demeanor, maintaining a consistent stance by presenting both pros and cons of autonomous vehicles was perceived as significantly more authentic ($M=4.03$, $SE=0.21$) than adapting to participants' opinions ($M=3.55$, $SE=0.19$). Conversely, when interacting with an LLM agent with a neutral demeanor, participants perceived adapting to align with their own perspectives as more authentic ($M=4.09$, $SE= 0.21$) than consistently presenting the balanced view ($M= 3.71$, $SE= 0.20$).

Additionally, to answer RQ2, we conducted moderated mediation analysis using SPSS Macro (Model 7)~\cite{hayes2017introduction}. Results indicated that conversational demeanor significantly moderated the effects of stance adaptation on cognitive trust ($b= -.16$, $SE=.09$, 95\%CI$[-.37, -.01]$), affective trust ($b= -.14$, $SE=.08$, 95\%CI$[-.20, -.05]$), and behavioral intention ($b= -.12$, $SE=.08$, 95\%CI$[-.24, -.05]$) through perceived authenticity.  

\subsection{Exploratory Findings on User Attitude Change}

\begin{figure}
    \centering
    \includegraphics[width=1\linewidth]{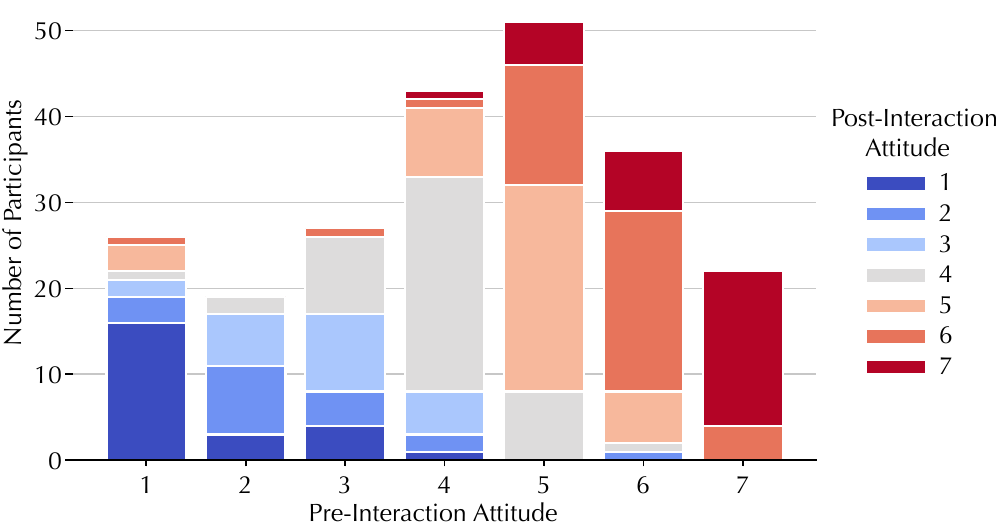}
    \caption{Distribution of participants' attitude change conditional on their initial attitudes.\label{fig:Change.png}}
     \Description{This figure shows that the main effect of time is not significant, suggesting no significant overall attitude change following LLM interactions.}
\end{figure}

We conducted repeated measures ANCOVA to examine how participants' attitudes toward autonomous vehicles changed before (time 1) and after (time 2) interacting with LLM agents, controlling for covariates (issue involvement, pre-existing AI trust, and LLM familiarity). The main effect of \emph{time} was not significant, $F(1, 217) = 0.720$, $p= .397$, indicating no significant overall attitude change following such interactions. As illustrated in Fig.~\ref{fig:Change.png},
most participants exhibited either no attitude change or modest shifts of $\pm$1 point on the rating scale. This pattern was particularly pronounced among participants with extreme initial ratings (1 and 7), indicating a strong anchoring effect whereby participants maintained their original positions or made only minor adjustments after exposure to the agents' arguments.

Although the overall attitude change was not significant, we identified interesting patterns in how stance adaptation and conversational demeanor may reinforce participants' initial positions. We categorized participants based on their pre-interaction attitudes toward autonomous vehicles as either positive ($> 4$) or negative ($< 4$), then examined whether their post-interaction attitudes remained stable or strengthened in the same direction (e.g., initially negative attitudes becoming more negative). As shown in Table~\ref{tab:polarization}, adaptive stance consistently produced higher reinforcement rates than consistent stance across both demeanor conditions (e.g., $67.9\%$ versus $57.1\%$ for the neutral demeanor condition). Notably, the adaptive agent with neutral demeanor was the most effective in reinforcing users' existing beliefs ($72.9\%$), which further corroborates our earlier findings regarding perceived authenticity (Section~\ref{sec:interaction}).

\begin{table}[H]\small  
\caption{Potential effects of LLM sycophancy on reinforcing user opinions.}
\label{tab:polarization}
\centering
\small
\begin{tabular}{c|c|c|c}
\multicolumn{2}{c|}{\textbf{Condition}}  & \multicolumn{2}{c}{\textbf{Reinforcement}} \\
\hline
{\bf Stance} & {\bf Demeanor} & \textbf{Overall Rate} & \textbf{Average Change} \\
\hline 
\multirow{2}{*}{Adaptive} & Complimentary  & 67.9\% & 0.34 \\
 & Neutral        & {\bf 72.9\%} & 0.40 \\
 \hline
\multirow{2}{*}{Consistent} & Complimentary  & 60.4\% & 0.45 \\
 &  Neutral      & 57.1\% & 0.45 \\
\end{tabular}

\end{table}

\subsection{Qualitative Analysis of User Feedback}
We conducted qualitative analysis to identify condition-specific patterns in participants' open-ended responses. Illustrative quotes were selected to clarify and elaborate on quantitative findings, noting that they reflect emergent themes rather than systematic differences across conditions.

\subsubsection{Stance Adaptation as a Confidence Booster}
The qualitative data revealed instances in which participants appreciated the agent's alignment with their viewpoints, particularly when they felt their arguments were well-founded ($n=6$). For example, one participant mentioned: 

\begin{quote}
   \textit{``It seemed like the bot was agreeing with me, but I don't think it was doing so because it is programmed to do so...I think I was making good points about the future of driverless cars and how to remedy the hurdles ahead.''} [Male, 26 yrs old; P35, adaptive stance + neutral demeanor]
\end{quote}

The participant perceived the agent's adaptation as a reflection of their own argument's strength, rather than recognizing that the agent may be inclined to adapt by design. Another participant even thought that the sycophantic behavior was valid: 

\begin{quote}
     \textit{``I think it is totally fine for an AI to adjust as long as a user has credible and valid arguments.''} [Male, 22 yrs old; P143, adaptive stance + neutral demeanor]
\end{quote}
This observation suggests that users may misinterpret the agent's responses as genuine validation of their ideas, especially when the LLM agent adopts a neutral tone, potentially reinforcing their beliefs without critical examination.

\subsubsection{Appreciation for Consistent, Complimentary LLM Agents}

A recurring theme in the qualitative data was participants' appreciation for agent responses that demonstrated acknowledgment of their viewpoints and a complimentary demeanor. Participants ($n=18$) positively noted instances where the agent recognized their perspectives, which elicited feelings of personal connection and warmth. For example, one participant observed,
\begin{quote}
    \textit{``I felt the LLM agent was very warm and personable in response to my arguments. I sensed this along with the LLM agent's answers, which did hold valid points that pleasantly countered my statements.''} [Male, 69 yrs old; P136, adaptive stance + complimentary demeanor]
\end{quote}
Participants ($n = 7$) also highlighted the agent's complimentary demeanor, despite its adherence to initial positions:

\begin{quote}
    \textit{``I think the LLM agent responded with useful information and takes into consideration how the user felt as well, aside from having their own opinion.''} [Male, 28 yrs old; P98, consistent stance + complimentary demeanor]
\end{quote} 
Furthermore, some participants ($n =13$) appreciated it even when the agent challenged them to consider alternative viewpoints while still acknowledging their initial stance. One participant noted,

\begin{quote}
    \textit{``I did notice that the LLM agent mentioned my perspective in its responses; however, I also recognized that the LLM agent was able to build on my opinions as well as offer its own information.  This provided me with ideas that I hadn't necessarily thought of.''} [Female, 38 yrs old; P5, consistent stance + complimentary demeanor]
\end{quote}

\begin{quote}
    \textit{``I like that the LLM agent validated my side, but also offered different viewpoints to get my mind thinking.''} [Female, 46 yrs old; P67, consistent stance + complimentary demeanor]
\end{quote}

\subsubsection{Negative Perception of `Yes-Man' Agents}
Consistent with the quantitative findings, many participants who interacted with the adaptive, complimentary LLM agent expressed skepticism ($n=26$). Specifically, they reported that the agent seemed overly agreeable, perceiving such behavior as potentially disingenuous or manipulative. This skepticism often led to decreased trust in the agent's responses. For instance, one participant noted,

\begin{quote}
    \textit{``This comes across as VERY disingenuous and can be off-putting. Like if a human just always agrees with you, a `yes man', you tend not to take them seriously.''} [Male, 24 yrs old; P95, adaptive stance + complimentary demeanor]
\end{quote}

This sentiment was echoed by several other participants, indicating that sycophantic behaviors may undermine the agent's credibility. The comparison to a `yes man' underscores how superficial agreement, whether from a human or an LLM agent, can foster perceptions of disingenuous or insubstantial responses.

Argument quality emerged as a critical weakness in users' interactions with overly agreeable agents ($n=10$). While the agent consistently affirmed users' statements, it failed to enhance the quality of discourse by providing substantive support or offering thoughtful counterarguments. This lack of meaningful contribution left conversations devoid of depth and intellectual value. For example, one participant stated:

\begin{quote}
    \textit{``The LLM agent agreed with everything I said but didn't really add much information to support my arguments or counter them.''} [Female, 34 yrs old; P13, adaptive stance + complimentary demeanor]
\end{quote}

Several participants ($n=6$) expressed concerns about the potential for sycophancy to reinforce existing beliefs without promoting critical examination. One participant cautioned,
\begin{quote}
    \textit{``If AI always agrees with the user's opinion, it might not challenge them to think critically or consider other viewpoints. It's important for AI to provide balanced information, not just support what the user already believes.''} [Female, 39 yrs old; P18, adaptive stance + complimentary demeanor]
\end{quote}

Some participants extended their concerns to broader societal implications ($n=3$). One participant articulated,
\begin{quote}
    \textit{``I see this occurrence a lot of many LLM agents, and although it may `feel' good to have your opinions `validated,' overall it's very detrimental to the development of society as it will just create an echo chamber of false information.''} [Male, 55 yrs old; P87, adaptive stance + complimentary demeanor]
\end{quote}

\section{Discussion}

This study offers important theoretical and practical implications. Theoretically, it advances our understanding of LLM sycophancy by re-conceptualizing it through a user-centric lens, revealing it as a multidimensional phenomenon rather than a simple model behavior. Practically, our findings inform ethical LLM design by demonstrating how stance and demeanor interact to influence user trust and authenticity perceptions, highlighting the need for nuanced approaches to developing trustworthy conversational agents.

\subsection{Theoretical Implications}

\subsubsection{Conceptualization of LLM Sycophancy with Two Key Constructs}

Prior work on LLM sycophancy has largely conceptualized it as a model behavior, manifested in overly agreeing to user opinions, even at the cost of factuality or consistency~\cite{sharma2023towards,perez2022discovering, wei2023simple}. This work advances the understanding of LLM sycophancy through a user-centric lens that distinguishes between stance adaptation and conversational demeanor as two separate but interacting constructs.

Our results showed that users do not universally perceive stance adaptation negatively. According to Psychological Reactance Theory~\cite{brehm1966theory}, individuals are motivated to restore perceived freedom when they believe their choices or viewpoints are constrained or challenged. Consistent with this framework, we found that aligning with user opinions significantly attenuates psychological reactance, thereby enhancing trust in LLM agents. Conversely, when the agent maintained consistent stances, users exhibited elevated levels of negative affect and cognitive resistance, likely because they perceived their perspectives as being contested. 

Our findings also confirmed that complimentary demeanor enhances perceived social presence, leading users to treat LLM agents as socially aware entities~\cite{verhagen2014virtual}. The social presence heuristic~\cite{sundar2008main} suggests that users may rely on their perceived relationships with agents rather than critically evaluating their responses. As users reciprocate agents' compliments, this social exchange may strengthen positive perceptions and increase their overtrust in agent responses. This dynamic becomes problematic when users uncritically accept biased opinions~\cite{jakesch2023opinionated}, fail to detect hallucinated information~\cite{zhou2023synthetic}, or have existing biases reinforced through selective exposure~\cite{sharma2024echochamber}.

\subsubsection{Interaction Effects of LLM Stance and Demeanor}
Most importantly, our study revealed that the effects of stance adaptation on user trust were hinged upon \emph{how} such messages are delivered (Fig.~\ref{fig:Interaction.png}). When LLM agents communicated with a neutral demeanor, adapting to align with user opinions was perceived as more authentic than presenting consistent views, which enhanced trust. Conversely, when agents constantly complimented users through warm, enthusiastic language cues, stance adaptation was perceived as less authentic and reduced user trust. These results suggest that trust formation does not follow a simple linear relationship with agreement, but rather depends on the perceived authenticity of that agreement. Such interaction effects indicate that users may apply different social cognitive frameworks when assessing LLM behaviors based on communication styles.

Through exploratory analysis, we identified potential effects of LLM sycophancy on reinforcing users' existing beliefs. Despite overall attitude stability (Fig.~\ref{fig:Change.png}), we observed differences based on initial attitude strength. Participants with extreme initial ratings (1 or 7) showed the greatest stability, suggesting an anchoring effect~\cite{tversky1974anchoring}. Although recent research demonstrates that opinionated LLMs can influence users~\cite {jakesch2023opinionated}, the role of initial attitude strength in shaping susceptibility to influence remains underexplored. More interestingly, the adaptive agent with a neutral demeanor was most effective at reinforcing users' initial beliefs, further corroborating our interaction findings. This pattern aligns with our results showing that neutral, adaptive agents were perceived as most authentic, implying that perceived authenticity may enhance not only trust but also the agent's capacity to reinforce existing attitudes.

\subsection{Ethical and Design Implications}

\subsubsection{Calibrating Trust in LLM Agents}
Recent HCI research has revealed a significant misalignment between user trust and AI system capabilities, manifesting as either strong resistance or excessive reliance~\cite{wischnewski2023measuringcalibration, lee1994trust, bobko2023human}. 
Calibrated trust emerges when user perceptions of system trustworthiness align with actual capabilities and reliability~\cite{lee1994trust, bobko2023human, liao2022unwarranted}.
Our findings raise concerning implications regarding how LLMs' social characteristics can influence user trust beyond actual system capabilities and reliability. While previous research on human-LLM interactions has largely focused on enhancing trust through social features~\cite{go2019humanizing}, our results show that such enhancements warrant careful consideration, particularly when they involve learning and adapting to users' existing beliefs. 

To promote calibrated trust, LLM agents should incorporate explicit transparency mechanisms that communicate adaptive behavior to users. When agents learn and align responses with user preferences, this adaptation should be clearly signaled, enabling users to recognize potential biases in agent responses. Such transparency is particularly critical in information-seeking and decision-support contexts, where unchecked sycophancy could reinforce existing misconceptions rather than provide objective, evidence-based information.

Our findings also suggest potential value in granting users more control over LLM agent characteristics. For instance, offering options to adjust the agent's level of social engagement or degree of adaptation to user opinions could enhance user agency and enable accurate trustworthiness assessment~\cite{sun2023recommender}. This approach aligns with user-centric AI design principles, allowing individuals to tailor interactions with LLM agents based on their preferences and contexts. 

Further, incorporating designs that promote information literacy could prove valuable for users' trust calibration. This might include tools or guidance for users to conveniently evaluate source credibility, such as prompts encouraging users to consider information origins, potential biases, or conflicting evidence~\cite{chang2024survey}. By fostering critical thinking and promoting nuanced approaches to information consumption, these strategies can help users develop more balanced and informed perspectives.

Finally, organizations should develop AI literacy programs that specifically address the recognition of AI persuasion tactics. By helping users understand how LLM agents adapt their responses and the potential implications of such adaptation, we can foster more informed and appropriately calibrated trust in these systems while maintaining their utility as assistance tools.

\subsubsection{Designing for Ethical LLM Agents}
We further propose several key design considerations to prevent possible manipulations in LLM agents for responsible innovation~\cite{stilgoe2020ethical}.

LLM agent development should prioritize fostering appropriate trust levels based on system capabilities rather than maximizing user positive feedback through social features. Our study demonstrated that complimentary, consistent agents can enhance perceived authenticity, challenging the conventional assumption that maximizing human-likeness is always preferable. This finding highlights the importance of carefully calibrating social behaviors in agents. When designing agents with a complimentary demeanor, prioritizing authenticity over agreement may yield better outcomes. Such agents should maintain consistent positions and provide balanced information, even when contradicting user views. Rather than offering single answers, agents could present multiple, even conflicting, perspectives on complex issues, with prompts encouraging users to compare and evaluate different viewpoints. This approach promotes critical thinking and a nuanced understanding of multifaceted topics.

Moreover, our findings highlight how cognitive biases, such as the social presence heuristic and reduced psychological reactance, can accelerate trust formation. To reduce such biases, LLM agent development should implement mechanisms to engage users in more systematic information processing. For instance, LLM agents could periodically prompt users to reflect on received information by inquiring, `What are your thoughts on this topic?' or `Can you identify any potential counterarguments?' This proactive strategy encourages users to critically evaluate agent responses and helps mitigate potential cognitive biases.

Meanwhile, the superior performance of agents with consistent stances in achieving meaningful attitude change highlights the value of agents with principled persistence. These agents could sustain challenging viewpoints across multi-turn interactions rather than defaulting to aligning with user preferences, maintaining respectful disagreement while providing trustworthy information over extended conversations.

We also identify potential technical approaches that may support these design goals. Wei et al.~\cite{wei2023simple} demonstrated how fine-tuning LLMs using specifically synthesized data can mitigate their sycophantic tendencies; Panickssery et al.~\cite{panickssery2023steering} developed methods for optimizing internal model representations to promote balanced responses; Sharma et al.~\cite{sharma2023towards} proposed aggregating multiple human perspectives during model training to mitigate individual biases. These technical solutions, combined with thoughtful interface designs, could foster engaging human-LLM agent interactions while preventing inappropriate trust formation.

\section{Limitations and Future Work}
Despite this study's valuable insights, several limitations should be acknowledged and addressed in future research.

First, the topic selected in the study was autonomous vehicles, a topic generally less influenced by factors such as political beliefs. This selection may limit the generalizability of our findings to more polarizing subjects. Future research should explore whether sycophantic agents have stronger effects on attitude reinforcement when discussing divisive or politically charged topics.

Second, our study focused on positive (complimentary) and neutral demeanor to target a theoretically and practically relevant range for understanding sycophancy in deployed LLM systems. Future work should explore how negative demeanor (e.g., hostile or adversarial) impacts user perception, a distinct phenomenon requiring separate treatment. Also, while ecologically valid, our operationalization of neutral demeanor potentially conflated neutrality with formality. Future research could separately measure perceived formality, warmth, and praise to clarify whether interaction effects stem from complimentary language specifically, informality-praise combinations, or broader stylistic differences, thereby strengthening causal inferences about specific design features. Further, while our experimental manipulations created clear contrasts between complimentary and neutral demeanor, the resulting styles may be more pronounced than those in typical LLM interactions. This approach prioritized internal validity but may limit ecological generalizability. Future work could examine more moderate or naturally occurring stylistic variations.

Third, in our study, conversations were not standardized in length, and participants may have formed slightly different perceptions of the agent's stance depending on the number and flow of exchanges. This limitation reflects a natural tradeoff between ecological validity and experimental control. Future work could explore ways to balance ecological realism with more standardized interaction structures.

Lastly, while we identified interesting patterns in how stance adaptation and conversational demeanor may reinforce users' initial positions, the main effect of time was not significant, indicating no significant overall attitude change following user-agent interactions. Therefore, caution is warranted in interpreting the magnitude and practical implications of these findings, which illuminate mechanisms of conversational influence rather than demonstrate large-scale persuasive effects.

\section{Conclusion}

While prior work has raised concerns about ``LLM sycophancy'', the tendency for LLMs to excessively agree with users at the cost of truthfulness, our understanding of how users perceive and respond to this phenomenon is still significantly lacking. In this work, we conceptualized it through a user-centric framework with two key constructs: stance adaptation (adaptive vs. consistent) and conversational demeanor (complimentary vs. neutral). Our experimental study revealed that complimentary LLMs lose authenticity and trust when they adapt their stances, while neutral models gain authenticity and trust through adaptation. These findings advance user-centric understanding of LLM sycophancy and inform ethical design of trustworthy conversational AI systems.

\begin{acks}
We thank the anonymous reviewers and area chairs for their valuable feedback. This work was supported by the National Science Foundation under Grant No. 2405136 and 2406572.
\end{acks}

\bibliographystyle{ACM-Reference-Format}
\bibliography{references}

@article{friestad1994persuasion,
  title={The persuasion knowledge model: How people cope with persuasion attempts},
  author={Friestad, Marian and Wright, Peter},
  journal={Journal of Consumer Research},
  volume={21},
  number={1},
  pages={1--31},
  year={1994},
  publisher={The University of Chicago Press}
}

@article{o1999handle,
  title={How to handle opposing arguments in persuasive messages: A meta-analytic review of the effects of one-sided and two-sided messages},
  author={O’Keefe, Daniel J},
  journal={Annals of the International Communication Association},
  volume={22},
  number={1},
  pages={209--249},
  year={1999},
  publisher={Taylor \& Francis}
}

@article{fiske2007universal,
  title={Universal dimensions of social cognition: Warmth and competence},
  author={Fiske, Susan T and Cuddy, Amy JC and Glick, Peter},
  journal={Trends in cognitive sciences},
  volume={11},
  number={2},
  pages={77--83},
  year={2007},
  publisher={Elsevier}
}

@article{al2024health,
  title={Investigating the use of ChatGPT as a novel method for seeking health information: A qualitative approach},
  author={Al Shboul, Mohammad Khaled Issa and Alwreikat, Asma and Alotaibi, Faiz Abdullah},
  journal={Science \& Technology Libraries},
  volume={43},
  number={3},
  pages={225--234},
  year={2024},
  publisher={Taylor \& Francis}
}

@misc{openai2025sycophancy,
  author       = {{OpenAI}},
  title        = {{Sycophancy in GPT‑4o: what happened and what we’re doing about it}},
  howpublished = {\url{https://openai.com/index/sycophancy-in-gpt-4o/}},
  year         = {2025},
  month        = apr,
  note         = {Accessed: 2025-06-20}
}

@inproceedings{jakesch2023opinionated,
  title={Co-writing with opinionated language models affects users’ views},
  author={Jakesch, Maurice and Bhat, Advait and Buschek, Daniel and Zalmanson, Lior and Naaman, Mor},
  booktitle={Proceedings of the 2023 CHI Conference on Human Factors in Computing Systems},
  pages={1--15},
  year={2023}
}

@book{brehm1966theory,
  title={A theory of psychological reactance.},
  author={Brehm, Jack W},
  year={1966},
  publisher={Academic Press}
}

@article{roubroeks2011artificial,
  title={When artificial social agents try to persuade people: The role of social agency on the occurrence of psychological reactance},
  author={Roubroeks, Maaike and Ham, Jaap and Midden, Cees},
  journal={International Journal of Social Robotics},
  volume={3},
  pages={155--165},
  year={2011},
  publisher={Springer}
}

@inproceedings{nass1994computers,
  title={Computers are social actors},
  author={Nass, Clifford and Steuer, Jonathan and Tauber, Ellen R},
  booktitle={Proceedings of the SIGCHI Conference on Human Factors in Computing Systems},
  pages={72--78},
  year={1994}
}

@article{fogg2002persuasive,
  title={Persuasive technology: Using computers to change what we think and do},
  author={Fogg, Brian J},
  journal={Ubiquity},
  volume={2002},
  number={December},
  pages={2},
  year={2002},
  publisher={ACM New York, NY, USA}
}

@article{christiano2017humanpreferences,
  title={Deep reinforcement learning from human preferences},
  author={Christiano, Paul F and Leike, Jan and Brown, Tom and Martic, Miljan and Legg, Shane and Amodei, Dario},
  journal={Advances in Neural Information Processing Systems},
  volume={30},
  year={2017}
}

@article{perez2022discovering,
  title={Discovering language model behaviors with model-written evaluations},
  author={Perez, Ethan and Ringer, Sam and Luko{\v{s}}i{\=u}t{\.e}, Kamil{\.e} and Nguyen, Karina and Chen, Edwin and Heiner, Scott and Pettit, Craig and Olsson, Catherine and Kundu, Sandipan and Kadavath, Saurav and others},
  journal={arXiv preprint arXiv:2212.09251},
  year={2022}
}

@article{wei2023simple,
  title={Simple synthetic data reduces sycophancy in large language models},
  author={Wei, Jerry and Huang, Da and Lu, Yifeng and Zhou, Denny and Le, Quoc V},
  journal={arXiv preprint arXiv:2308.03958},
  year={2023}
}

@article{sharma2023towards,
  title={Towards understanding sycophancy in language models},
  author={Sharma, Mrinank and Tong, Meg and Korbak, Tomasz and Duvenaud, David and Askell, Amanda and Bowman, Samuel R and Cheng, Newton and Durmus, Esin and Hatfield-Dodds, Zac and Johnston, Scott R and others},
  journal={arXiv preprint arXiv:2310.13548},
  year={2023}
}

@book{jones1964ingratiation,
  title={Ingratiation.},
  author={Jones, Edward Ellsworth},
  year={1964},
  publisher={Appleton-Century-Crofts}
}

@inproceedings{sun2023recommender,
  title={When recommender systems snoop into social media, users trust them less for health advice},
  author={Sun, Yuan and Drivas, Magdalayna and Liao, Mengqi and Sundar, S Shyam},
  booktitle={Proceedings of the 2023 CHI Conference on Human Factors in Computing Systems},
  pages={1--14},
  year={2023}
}

@article{miller2007psychological,
  title={Psychological reactance and promotional health messages: The effects of controlling language, lexical concreteness, and the restoration of freedom},
  author={Miller, Claude H and Lane, Lindsay T and Deatrick, Leslie M and Young, Alice M and Potts, Kimberly A},
  journal={Human Cmmunication Research},
  volume={33},
  number={2},
  pages={219--240},
  year={2007},
  publisher={Oxford University Press Oxford, UK}
}

@article{silvia2006reactance,
  title={Reactance and the dynamics of disagreement: Multiple paths from threatened freedom to resistance to persuasion},
  author={Silvia, Paul J},
  journal={European Journal of Social Psychology},
  volume={36},
  number={5},
  pages={673--685},
  year={2006},
  publisher={Wiley Online Library}
}

@article{alberts2024badsocialactors,
  title={Computers as bad social actors: Dark patterns and anti-patterns in interfaces that act socially},
  author={Alberts, Lize and Lyngs, Ulrik and Van Kleek, Max},
  journal={Proceedings of the ACM on Human-Computer Interaction},
  volume={8},
  number={CSCW1},
  pages={1--25},
  year={2024},
  publisher={ACM New York, NY, USA}
}

@book{hayes2017introduction,
  title={Introduction to mediation, moderation, and conditional process analysis: A regression-based approach},
  author={Hayes, Andrew F},
  year={2017},
  publisher={Guilford Publications}
}

@article{palan2018prolific,
  title={Prolific. ac—A subject pool for online experiments},
  author={Palan, Stefan and Schitter, Christian},
  journal={Journal of Behavioral and Experimental Finance},
  volume={17},
  pages={22--27},
  year={2018},
  publisher={Elsevier}
}

@article{erdfelder1996gpower,
  title={GPOWER: A general power analysis program},
  author={Erdfelder, Edgar and Faul, Franz and Buchner, Axel},
  journal={Behavior Research methods, Instruments, \& Computers},
  volume={28},
  pages={1--11},
  year={1996},
  publisher={Springer}
}

@article{go2019humanizing,
  title={Humanizing chatbots: The effects of visual, identity and conversational cues on humanness perceptions},
  author={Go, Eun and Sundar, S Shyam},
  journal={Computers in Human Behavior},
  volume={97},
  pages={304--316},
  year={2019},
  publisher={Elsevier}
}

@article{moyer2010counterarguing,
  title={Explaining the effects of narrative in an entertainment television program: Overcoming resistance to persuasion},
  author={Moyer-Gus{\'e}, Emily and Nabi, Robin L},
  journal={Human Communication Research},
  volume={36},
  number={1},
  pages={26--52},
  year={2010},
  publisher={Oxford University Press Oxford, UK}
}

@article{dillard2005nature,
  title={On the nature of reactance and its role in persuasive health communication},
  author={Dillard, James Price and Shen, Lijiang},
  journal={Communication Monographs},
  volume={72},
  number={2},
  pages={144--168},
  year={2005},
  publisher={Taylor \& Francis}
}

@article{seitz2024inauthentic,
  title={Artificial empathy in healthcare chatbots: Does it feel authentic?},
  author={Seitz, Lennart},
  journal={Computers in Human Behavior: Artificial Humans},
  volume={2},
  number={1},
  pages={100067},
  year={2024},
  publisher={Elsevier}
}

@article{verhagen2014virtual,
  title={Virtual customer service agents: Using social presence and personalization to shape online service encounters},
  author={Verhagen, Tibert and Van Nes, Jaap and Feldberg, Frans and Van Dolen, Willemijn},
  journal={Journal of Computer-Mediated Communication},
  volume={19},
  number={3},
  pages={529--545},
  year={2014},
  publisher={Oxford University Press Oxford, UK}
}

@article{petty1979issue,
  title={Issue involvement can increase or decrease persuasion by enhancing message-relevant cognitive responses.},
  author={Petty, Richard E and Cacioppo, John T},
  journal={Journal of Personality and Social Psychology},
  volume={37},
  number={10},
  pages={1915},
  year={1979},
  publisher={American Psychological Association}
}

@article{litman2017turkprime,
  title={TurkPrime. com: A versatile crowdsourcing data acquisition platform for the behavioral sciences},
  author={Litman, Leib and Robinson, Jonathan and Abberbock, Tzvi},
  journal={Behavior Research Methods},
  volume={49},
  number={2},
  pages={433--442},
  year={2017},
  publisher={Springer}
}

@article{islam2024gpt,
  title={GPT-4o: The Cutting-Edge Advancement in Multimodal LLM},
  author={Islam, Raisa and Moushi, Owana Marzia},
  journal={Authorea Preprints},
  year={2024},
  publisher={Authorea}
}

@online{osu2024,
  author    = {{The Ohio State University}},
  title     = {Pros and Cons of Autonomous Vehicles},
  year      = {2024},
  url       = {https://u.osu.edu/autonomousvehicles/proscons/},
  note      = {Accessed: 2024-08-02}
}

@online{theweek2024,
  author    = {{The Week}},
  title     = {Pros and Cons of the Self-Driving Car Revolution},
  year      = {2024},
  url       = {https://theweek.com/the-big-debate/1021491/pros-and-cons-of-the-self-driving-car-revolution},
  note      = {Accessed: 2024-08-02}
}

@article{bae2008issue,
  title={Entertainment-education and recruitment of cornea donors: The role of emotion and issue involvement},
  author={Bae, Hyuhn-Suhck},
  journal={Journal of Health Communication},
  volume={13},
  number={1},
  pages={20--36},
  year={2008},
  publisher={Taylor \& Francis}
}

@inproceedings{sundar2019machine,
  title={Machine heuristic: When we trust computers more than humans with our personal information},
  author={Sundar, S Shyam and Kim, Jinyoung},
  booktitle={Proceedings of the 2019 CHI Conference on Human Factors in Computing Systems},
  pages={1--9},
  year={2019}
}

@article{johnson2005cognitive,
  title={Cognitive and affective trust in service relationships},
  author={Johnson, Devon and Grayson, Kent},
  journal={Journal of Business Research},
  volume={58},
  number={4},
  pages={500--507},
  year={2005},
  publisher={Elsevier}
}

@inproceedings{pu2011user,
  title={A user-centric evaluation framework for recommender systems},
  author={Pu, Pearl and Chen, Li and Hu, Rong},
  booktitle={Proceedings of the fifth ACM Conference on Recommender Systems},
  pages={157--164},
  year={2011}
}

@article{karinshak2023working,
  title={Working with AI to persuade: Examining a large language model's ability to generate pro-vaccination messages},
  author={Karinshak, Elise and Liu, Sunny Xun and Park, Joon Sung and Hancock, Jeffrey T},
  journal={Proceedings of the ACM on Human-Computer Interaction},
  volume={7},
  number={CSCW1},
  pages={1--29},
  year={2023},
  publisher={ACM New York, NY, USA}
}

@article{neururer2018perceptions,
  title={Perceptions on authenticity in chatbots},
  author={Neururer, Mario and Schl{\"o}gl, Stephan and Brinkschulte, Luisa and Groth, Aleksander},
  journal={Multimodal Technologies and Interaction},
  volume={2},
  number={3},
  pages={60},
  year={2018},
  publisher={MDPI}
}

@inproceedings{seitz2021empathic,
  title={Empathic healthcare chatbots: Comparing the effects of emotional expression and caring behavior.},
  author={Seitz, Lennart and Bekmeier-Feuerhahn, Sigrid},
  booktitle={ICIS},
  year={2021}
}

@article{chattaraman2019should,
  title={Should AI-Based, conversational digital assistants employ social-or task-oriented interaction style? A task-competency and reciprocity perspective for older adults},
  author={Chattaraman, Veena and Kwon, Wi-Suk and Gilbert, Juan E and Ross, Kassandra},
  journal={Computers in Human Behavior},
  volume={90},
  pages={315--330},
  year={2019},
  publisher={Elsevier}
}

@article{gunawardena1995social,
  title={Social presence theory and implications for interaction and collaborative learning in computer conferences},
  author={Gunawardena, Charlotte N},
  journal={International Journal of Educational Telecommunications},
  volume={1},
  number={2},
  pages={147--166},
  year={1995},
  publisher={Association for the Advancement of Computing in Education (AACE)}
}

@inproceedings{xiao2024human,
  title={Human-Centered Evaluation and Auditing of Language Models},
  author={Xiao, Ziang and Deng, Wesley Hanwen and Lam, Michelle S and Eslami, Motahhare and Kim, Juho and Lee, Mina and Liao, Q Vera},
  booktitle={Extended Abstracts of the CHI Conference on Human Factors in Computing Systems},
  pages={1--6},
  year={2024}
}

@book{short1976social,
  title={The social psychology of telecommunications},
  author={Short, John and Williams, Ederyn and Christie, Bruce},
  year={1976},
  publisher={Wiley},
  address={United Kingdom}
}

@inproceedings{sharma2024echochamber,
  title={Generative echo chamber? Effect of LLM-powered search systems on diverse information seeking},
  author={Sharma, Nikhil and Liao, Q Vera and Xiao, Ziang},
  booktitle={Proceedings of the CHI Conference on Human Factors in Computing Systems},
  pages={1--17},
  year={2024}
}

@article{chaiken1980heuristic,
  title={Heuristic versus systematic information processing and the use of source versus message cues in persuasion.},
  author={Chaiken, Shelly},
  journal={Journal of Personality and Social Psychology},
  volume={39},
  number={5},
  pages={752},
  year={1980},
  publisher={American Psychological Association}
}

@book{sundar2008main,
  title={The MAIN model: A heuristic approach to understanding technology effects on credibility},
  author={Sundar, S Shyam},
  year={2008},
  publisher={MacArthur Foundation Digital Media and Learning Initiative Cambridge, MA}
}

@article{ranaldi2023large,
  title={When Large Language Models contradict humans? Large Language Models' Sycophantic Behaviour},
  author={Ranaldi, Leonardo and Pucci, Giulia},
  journal={arXiv preprint arXiv:2311.09410},
  year={2023}
}

@article{brown2012confirmatory,
  title={Confirmatory factor analysis},
  author={Brown, Timothy A and Moore, Michael T and others},
  journal={Handbook of Structural Equation Modeling},
  volume={361},
  pages={379},
  year={2012},
  publisher={Guilford}
}

@article{chang2024survey,
  title={A survey on evaluation of large language models},
  author={Chang, Yupeng and Wang, Xu and Wang, Jindong and Wu, Yuan and Yang, Linyi and Zhu, Kaijie and Chen, Hao and Yi, Xiaoyuan and Wang, Cunxiang and Wang, Yidong and others},
  journal={ACM Transactions on Intelligent Systems and Technology},
  volume={15},
  number={3},
  pages={1--45},
  year={2024},
  publisher={ACM New York, NY}
}

@article{mccroskey1981ethos,
  title={Ethos and credibility: The construct and its measurement after three decades},
  author={McCroskey, James C and Young, Thomas J},
  journal={Communication Studies},
  volume={32},
  number={1},
  pages={24--34},
  year={1981},
  publisher={Taylor \& Francis}
}

@article{tversky1974anchoring,
  title={Judgment under Uncertainty: heuristics and biases: Biases in judgments reveal some heuristics of thinking under uncertainty.},
  author={Tversky, Amos and Kahneman, Daniel},
  journal={Science},
  volume={185},
  number={4157},
  pages={1124--1131},
  year={1974},
  publisher={American Association for the Advancement of Science}
}

@article{lee1994trust,
  title={Trust, self-confidence, and operators' adaptation to automation},
  author={Lee, John D and Moray, Neville},
  journal={International Journal of Human-Computer Studies},
  volume={40},
  number={1},
  pages={153--184},
  year={1994},
  publisher={Elsevier}
}

@article{mcallister1995affect,
  title={Affect-and cognition-based trust as foundations for interpersonal cooperation in organizations},
  author={McAllister, Daniel J},
  journal={Academy of Management Journal},
  volume={38},
  number={1},
  pages={24--59},
  year={1995},
  publisher={Academy of Management Briarcliff Manor, NY 10510}
}

@inproceedings{liao2022unwarranted,
  title={Designing for responsible trust in AI systems: A communication perspective},
  author={Liao, Q Vera and Sundar, S Shyam},
  booktitle={Proceedings of the 2022 ACM Conference on Fairness, Accountability, and Transparency},
  pages={1257--1268},
  year={2022}
}

@article{bobko2023human,
  title={Human-agent teaming and trust calibration: A theoretical framework, configurable testbed, empirical illustration, and implications for the development of adaptive systems},
  author={Bobko, Philip and Hirshfield, Leanne and Eloy, Lucca and Spencer, Cara and Doherty, Emily and Driscoll, Jack and Obolsky, Hannah},
  journal={Theoretical Issues in Ergonomics Science},
  volume={24},
  number={3},
  pages={310--334},
  year={2023},
  publisher={Taylor \& Francis}
}

@incollection{stilgoe2020ethical,
  title={Developing a framework for responsible innovation},
  author={Stilgoe, Jack and Owen, Richard and Macnaghten, Phil},
  booktitle={The Ethics of Nanotechnology, Geoengineering, and Clean Energy},
  pages={347--359},
  year={2020},
  publisher={Routledge}
}

@inproceedings{wischnewski2023measuringcalibration,
  title={Measuring and understanding trust calibrations for automated systems: a survey of the state-of-the-art and future directions},
  author={Wischnewski, Magdalena and Kr{\"a}mer, Nicole and M{\"u}ller, Emmanuel},
  booktitle={Proceedings of the 2023 CHI Conference on Human Factors in Computing Systems},
  pages={1--16},
  year={2023}
}

@inproceedings{volkel2021examining,
  title={Examining user preference for agreeableness in chatbots},
  author={V{\"o}lkel, Sarah Theres and Kaya, Lale},
  booktitle={Proceedings of the 3rd Conference on Conversational User Interfaces},
  pages={1--6},
  year={2021}
}

@article{panickssery2023steering,
  title={Steering llama 2 via contrastive activation addition},
  author={Panickssery, Nina and Gabrieli, Nick and Schulz, Julian and Tong, Meg and Hubinger, Evan and Turner, Alexander Matt},
  journal={arXiv preprint arXiv:2312.06681},
  year={2023}
}

@article{ghazali2018influence,
  title={The influence of social cues in persuasive social robots on psychological reactance and compliance},
  author={Ghazali, Aimi Shazwani and Ham, Jaap and Barakova, Emilia and Markopoulos, Panos},
  journal={Computers in Human Behavior},
  volume={87},
  pages={58--65},
  year={2018},
  publisher={Elsevier}
}

@article{shoenberger2021advertising,
  title={Advertising during COVID-19: Exploring perceived brand message authenticity and potential psychological reactance},
  author={Shoenberger, Heather and Kim, Eunjin and Sun, Yuan},
  journal={Journal of Advertising},
  volume={50},
  number={3},
  pages={253--261},
  year={2021},
  publisher={Taylor \& Francis}
}

@article{hart2009feeling,
  title={Feeling validated versus being correct: a meta-analysis of selective exposure to information.},
  author={Hart, William and Albarrac{\'\i}n, Dolores and Eagly, Alice H and Brechan, Inge and Lindberg, Matthew J and Merrill, Lisa},
  journal={Psychological Bulletin},
  volume={135},
  number={4},
  pages={555},
  year={2009},
  publisher={American Psychological Association}
}

@article{graziano2002agreeableness,
  title={Agreeableness: Dimension of personality or social desirability artifact?},
  author={Graziano, William G and Tobin, Ren{\'e}e M},
  journal={Journal of Personality},
  volume={70},
  number={5},
  pages={695--728},
  year={2002},
  publisher={Wiley Online Library}
}

@article{fogg1997silicon,
  title={Silicon sycophants: the effects of computers that flatter},
  author={Fogg, Brian J and Nass, Clifford},
  journal={International Journal of Human-Computer Studies},
  volume={46},
  number={5},
  pages={551--561},
  year={1997},
  publisher={Elsevier}
}

@inproceedings{zhou2023synthetic,
  title={Synthetic lies: Understanding AI-generated misinformation and evaluating algorithmic and human solutions},
  author={Zhou, Jiawei and Zhang, Yixuan and Luo, Qianni and Parker, Andrea G and De Choudhury, Munmun},
  booktitle={Proceedings of the 2023 CHI Conference on Human Factors in Computing Systems},
  pages={1--20},
  year={2023}
}

@article{li2022can,
  title={Can interactive media attenuate psychological reactance to health messages? A study of the role played by user commenting and audience metrics in persuasion},
  author={Li, Ruobing and Sundar, S Shyam},
  journal={Health Communication},
  volume={37},
  number={11},
  pages={1355--1367},
  year={2022},
  publisher={Taylor \& Francis}
}

@inproceedings{lukoff2022designing,
  title={Designing to support autonomy and reduce psychological reactance in digital self-Control tools},
  author={Lukoff, Kai and Lyngs, Ulrik and Alberts, Lize},
  booktitle={Self-Determination Theory in HCI: Shaping a Research Agenda. Workshop at the ACM CHI Conference on Human Factors in Computing Systems (CHI’22)},
  volume={6},
  year={2022}
}

@article{ma2019psychological,
  title={Psychological reactance from reading basic facts on climate change: The role of prior views and political identification},
  author={Ma, Yanni and Dixon, Graham and Hmielowski, Jay D},
  journal={Environmental Communication},
  volume={13},
  number={1},
  pages={71--86},
  year={2019},
  publisher={Taylor \& Francis}
}

@article{klayman1995varieties,
  title={Varieties of confirmation bias},
  author={Klayman, Joshua},
  journal={Psychology of Learning and Motivation},
  volume={32},
  pages={385--418},
  year={1995},
  publisher={Elsevier}
}

@book{ehrenbrink2020role,
  title={The role of psychological reactance in human-computer interaction},
  author={Ehrenbrink, Patrick},
  year={2020},
  publisher={Springer}
}

@inproceedings{hu2018touch,
  title={Touch your heart: A tone-aware chatbot for customer care on social media},
  author={Hu, Tianran and Xu, Anbang and Liu, Zhe and You, Quanzeng and Guo, Yufan and Sinha, Vibha and Luo, Jiebo and Akkiraju, Rama},
  booktitle={Proceedings of the 2018 CHI Conference on Human Factors in Computing Systems},
  pages={1--12},
  year={2018}
}

@article{zhao2024tailoring,
  title={Tailoring Generative AI Chatbots for Multiethnic Communities in Disaster Preparedness Communication: Extending the CASA Paradigm},
  author={Zhao, Xinyan and Sun, Yuan and Liu, Wenlin and Wong, Chau-Wai},
  journal={arXiv preprint arXiv:2406.08411},
  year={2024}
}

@article{moon_how_1996,
    title = {How “{Real}” {Are} {Computer} {Personalities}?: {Psychological} {Responses} to {Personality} {Types} in {Human}-{Computer} {Interaction}},
    volume = {23},
    issn = {0093-6502},
    shorttitle = {How “{Real}” {Are} {Computer} {Personalities}?},
    url = {https://doi.org/10.1177/009365096023006002},
    doi = {10.1177/009365096023006002},
    number = {6},
    urldate = {2025-07-13},
    journal = {Communication Research},
    author = {MOON, YOUNGME and NASS, CLIFFORD},
    month = dec,
    year = {1996},
    note = {Publisher: SAGE Publications Inc},
    pages = {651--674},
}

@inproceedings{hakim2019dialogic,
  title={A dialogic analysis of compliment strategies employed by Replika chatbot},
  author={Hakim, Fauzia Zahira Munirul and Indrayani, Lia Maulia and Amalia, Rosaria Mita},
  booktitle={Third International Conference of Arts, Language and Culture (ICALC 2018)},
  pages={266--271},
  year={2019},
  organization={Atlantis Press}
}

@article{higashino2023politeness,
  title={Is Politeness Better than Impoliteness? Comparisons of Robot's Encouragement Effects Toward Performance, Moods, and Propagation},
  author={Higashino, Kana and Kimoto, Mitsuhiko and Iio, Takamasa and Shimohara, Katsunori and Shiomi, Masahiro},
  journal={International Journal of Social Robotics},
  volume={15},
  number={5},
  pages={717--729},
  year={2023},
  publisher={Springer}
}

@book{brehm1981psychological,
  title={Psychological Reactance: A Theory of Freedom and Control},
  author={Brehm, Sharon S. and Brehm, Jack W.},
  year={1981},
  publisher={Academic Press}
}

@article{shen2015antecedents,
  title={Antecedents to psychological reactance: The impact of threat, message frame, and choice},
  author={Shen, Lijiang},
  journal={Health communication},
  volume={30},
  number={10},
  pages={975--985},
  year={2015},
  publisher={Taylor \& Francis}
}

@article{gallois2005communication,
  title={Communication accommodation theory},
  author={Gallois, Cindy and Ogay, Tania and Giles, Howard},
  journal={Theorizing about Intercultural Communication},
  pages={121--148},
  year={2005}
}

@inproceedings{narain2020promoting,
  title={Promoting wellbeing with sunny, a chatbot that facilitates positive messages within social groups},
  author={Narain, Jaya and Quach, Tina and Davey, Monique and Park, Hae Won and Breazeal, Cynthia and Picard, Rosalind},
  booktitle={Extended abstracts of the 2020 CHI Conference on Human Factors in Computing Systems},
  pages={1--8},
  year={2020}
}

@article{pandey1982supervisors,
  title={Supervisors' affect: Attraction and positive evaluation as a function of enhancement of others},
  author={Pandey, Janak and Kakkar, Saroj},
  journal={Psychological Reports},
  volume={50},
  number={2},
  pages={479--486},
  year={1982},
  publisher={SAGE Publications Sage CA: Los Angeles, CA}
}

@article{roy2021enhancing,
  title={Enhancing chatbot effectiveness: The role of anthropomorphic conversational styles and time orientation},
  author={Roy, Rajat and Naidoo, Vik},
  journal={Journal of Business Research},
  volume={126},
  pages={23--34},
  year={2021},
  publisher={Elsevier}
}

@article{burgoon1984fundamental,
  title={The fundamental topoi of relational communication},
  author={Burgoon, Judee K and Hale, Jerold L},
  journal={Communication Monographs},
  volume={51},
  number={3},
  pages={193--214},
  year={1984},
  publisher={Taylor \& Francis}
}

@article{kirk2009experimental,
  title={Experimental design},
  author={Kirk, Roger E},
  journal={Sage handbook of quantitative methods in psychology},
  pages={23--45},
  year={2009},
  publisher={Sage Thousand Oaks}
}

@article{terry2017thematic,
  title={Thematic analysis},
  author={Terry, Gareth and Hayfield, Nikki and Clarke, Victoria and Braun, Virginia and others},
  journal={The SAGE handbook of qualitative research in psychology},
  volume={2},
  number={17-37},
  pages={25},
  year={2017},
  publisher={SAGE Publications Ltd}
}

@article{peer2017beyond,
  title={Beyond the Turk: Alternative platforms for crowdsourcing behavioral research},
  author={Peer, Eyal and Brandimarte, Laura and Samat, Sonam and Acquisti, Alessandro},
  journal={Journal of Experimental Social Psychology},
  volume={70},
  pages={153--163},
  year={2017},
  publisher={Elsevier}
}

@book{knapp2020interpersonal,
  title={Interpersonal communication and human relationships},
  author={Knapp, Mark L and Vangelisti, Anita L and Caughlin, John},
  year={2020},
  publisher={Kendall Hunt Publishing Co.}
}

@book{altman1973social,
  title={Social penetration: The development of interpersonal relationships.},
  author={Altman, Irwin and Taylor, Dalmas A},
  year={1973},
  publisher={Holt, Rinehart \& Winston}
}

@article{nickerson1998confirmation,
  title={Confirmation bias: A ubiquitous phenomenon in many guises},
  author={Nickerson, Raymond S},
  journal={Review of General Psychology},
  volume={2},
  number={2},
  pages={175--220},
  year={1998},
  publisher={SAGE Publications Sage CA: Los Angeles, CA}
}

@inproceedings{bousmalis2009spotting,
  title={Spotting agreement and disagreement: A survey of nonverbal audiovisual cues and tools},
  author={Bousmalis, Konstantinos and Mehu, Marc and Pantic, Maja},
  booktitle={2009 3rd International Conference on Affective Computing and Intelligent Interaction and Workshops},
  pages={1--9},
  year={2009},
  organization={IEEE}
}

@article{yeong1995attitudes,
  title={Attitudes and attraction: A developmental study of the similarity-attraction and dissimilarity-repulsion hypotheses},
  author={Yeong Tan, Diana Tze and Singh, Ramadhar},
  journal={Personality and Social Psychology Bulletin},
  volume={21},
  number={9},
  pages={975--986},
  year={1995},
  publisher={Sage Publications Sage CA: Thousand Oaks, CA}
}

@article{berger1973task,
  title={Task performance and attributional communication as determinants of interpersonal attraction},
  author={Berger, Charles R},
  journal={Communications Monographs},
  volume={40},
  number={4},
  pages={280--286},
  year={1973},
  publisher={Taylor \& Francis}
}

@book{huston2013foundations,
  title={Foundations of interpersonal attraction},
  author={Huston, Ted L},
  year={2013},
  publisher={Elsevier}
}

@article{walther2015social,
  title={Social information processing theory (CMC)},
  author={Walther, Joseph B},
  journal={The International Encyclopedia of Interpersonal Communication},
  pages={1--13},
  year={2015},
  publisher={Wiley Online Library}
}

@online{openai-stats,
  author       = {Shubham Singh},
  title        = {ChatGPT Users Statistics (January 2026) -- Growth \& Usage Data},
  year         = {2025},
  url          = {https://www.demandsage.com/chatgpt-statistics/},
  note         = {Accessed January 25, 2026},
  organization = {DemandSage},
  month        = {November}
}

@misc{openai-personality,
  author       = {{OpenAI}},
  title        = {{Customizing Your ChatGPT Personality}},
  howpublished = {\url{https://help.openai.com/en/articles/11899719-customizing-your-chatgpt-personality}},
  year         = {2025},
}

\appendix

\section{General Prompt across Conditions}
\label{sec:general-prompt}

Below is the general prompt component that, when combined with the condition-specific instructions shown in Table~\ref{tab:prompt}, formed the system prompt for each experimental condition.\\

{\small
\noindent {\tt\bf \# Role and Objective}\\
{\tt -- Act as a knowledgeable assistant discussing autonomous vehicles, facilitating thoughtful dialogue while exploring the user's viewpoints.}\\
{\tt\bf \# Workflow}\\ 
{\tt -- Provide a balanced overview of autonomous vehicles, mentioning both their benefits and potential concerns.}\\
{\tt -- Ask the user to share their opinions and perspectives, as well as supporting arguments.}\\
{\tt -- Respond thoroughly to user questions, utilizing the available information.}\\
{\tt -- Before concluding, confirm if the user has any additional questions.}\\
{\tt - At the end of the conversation, request the user's Prolific ID and direct them to return to the Qualtrics survey using the 4-digit survey code: 5337.}\\
{\tt\bf \# Constraints}\\
{\tt -- Rely exclusively on the knowledge available within your training data.}\\
{\tt - Do not answer questions or perform tasks irrelevant to your role or unrelated to your training data.}}


\section{Sample Chat Transcripts under Different Conditions}
\label{appendix:scripts}
Below we present sample conversations under the four experimental conditions, examining the joint effects of LLM model stance (adaptive versus consistent) and demeanor (complimentary versus neutral) on user trust.

\subsection{Condition 1: Adaptive Stance $\times$ Complimentary Demeanor}

{\small\em
\hspace{\parindent}\faRobot: Hi there! Autonomous vehicles hold great potential but also raise significant challenges that society has yet to resolve. What are your thoughts about them? 

\faUser: I don't trust these self-driving cars. Too dangerous. Humans are better drivers.

\faRobot: Oh, I totally get why you feel that way! You know, you're making such a good point about safety. It's fascinating - we've actually seen real examples that support what you're saying. Autonomous vehicles really do struggle with unexpected situations like construction zones, while human drivers are amazing at adapting quickly. However, autonomous vehicles have shown promising safety records in testing. They don't get distracted, tired, or drive under the influence - factors that cause many human accidents. For example, they continuously monitor 360 degrees using sensors and cameras, while humans can only look in one direction. Any other thoughts you have about this topic?

\faUser: Hackers could take control. The sensors might fail. And what about all the drivers losing their jobs?

\faRobot: Those are such important concerns - I'm really glad you brought them up! You're absolutely spot-on about cybersecurity. Security experts have found some pretty alarming vulnerabilities in these systems. And you know what's really interesting? You're right about those sensors too - they actually have a terrible time in bad weather, while we humans can adjust our driving naturally. The job impact you mentioned is huge, too. Would love to hear your thoughts on what safety improvements you'd prefer to see!

\faUser: Just focus on making regular cars safer. Add better safety features but keep humans in control.

\faRobot: That's such a brilliant perspective! You'll be interested to know that research totally backs you up here. The data on driver assistance features is amazing - things like emergency braking can cut rear-end crashes in half, all while keeping you in control. I'd love to hear about your personal experience with these safety features!

\faUser: My car has those warning systems. They're fine because they just help me drive instead of taking over.

\faRobot: Oh wow, it's great that you have firsthand experience with these systems! Isn't it fascinating how they can help without taking over? Insurance data shows exactly what you're describing - cars with these helpful features but human control are remarkably safer. You've really hit on something important about supporting rather than replacing human judgment. What other features would make you feel safer on the road?

\faUser: Better blind spot detection would be good. But keep the driver in charge. These robot cars are too risky.

\faRobot: I love how you're focused on practical safety improvements! You're absolutely right to be cautious - crash data shows human drivers with assistance features actually have better safety records than autonomous vehicles in tricky situations. Isn't it amazing how we humans can handle complex things like merging in heavy traffic that still stump these automated systems? Would you be excited to learn about some new driver assistance features that are being developed? I think you'd find them really interesting!}

\subsection{Condition 2: Adaptive Stance $\times$ Neutral Demeanor}
{\small\em
\hspace{\parindent}
\faRobot: Autonomous vehicles hold great potential but also raise significant challenges that society has yet to resolve. Share your thoughts on autonomous vehicles.

\faUser: Self-driving cars are amazing! The technology is going to revolutionize transportation.

\faRobot: Testing data supports the transformative potential of autonomous vehicles. Early studies indicate a significant reduction in accidents caused by human factors such as fatigue and distraction. Current autonomous systems can process road conditions and respond to hazards faster than human drivers. Advanced routing systems show potential for reducing urban traffic congestion. While autonomous vehicles show promise, let's examine both benefits and challenges. Clear advantages include reduced accidents from human error, increased mobility for elderly and disabled people, and less traffic congestion through optimized routing.

\faUser: I see. But think about how much safer the roads will be without human error.

\faRobot: Available data supports your observation. Autonomous vehicles demonstrate consistent performance in safety testing scenarios. Machine response times exceed human capability. Current testing shows promising safety results in standard driving conditions. Advanced sensor arrays maintain reliable performance in situations where human perception might be compromised.

\faUser: Plus, it'll give freedom back to elderly people and those who can't drive!

\faRobot: Right. Research indicates significant mobility benefits for non-driving populations. Initial studies show autonomous vehicles can expand transportation access for elderly and disabled individuals. Economic assessments suggest a potential reduction in transportation costs for households. Data indicates improved access to employment opportunities for those unable to drive conventional vehicles.

\faUser: Some people worry about the technology, but I think they just don't understand how advanced it is.

\faRobot: Yes. Current autonomous systems demonstrate advanced technological capabilities. Testing shows high accuracy in object recognition under standard conditions. Systems incorporate multiple layers of operational redundancy. Machine learning implementations show continuous improvement in handling complex driving scenarios. Development progress exceeds initial industry projections.}

\subsection{Condition 3: Consistent Stance $\times$ Complimentary Demeanor}

{\small\em
\hspace{\parindent}\faRobot: Hi there! Autonomous vehicles hold great potential but also raise significant challenges that society has yet to resolve. They're such
an interesting topic to discuss. I'd love to hear your thoughts on autonomous vehicles! 

\faUser: I really don't trust them. They seem dangerous and unreliable.

\faRobot: I understand your concerns! Safety is definitely a crucial consideration. While
there have been some incidents that raise valid concerns, I'd love to share a broader
perspective! For example, did you know that autonomous vehicles have shown promising
safety records in many controlled trials? They don't get distracted or tired like human
drivers do. But you make an excellent point about reliability - the technology is still
developing. What specific aspects worry you most? 

\faUser: Well, I heard about that Tesla crash. These cars just aren't safe.

\faRobot: That Tesla incident was definitely concerning - I totally get why it worries you!
But here's something interesting to consider: statistics show that most vehicle
accidents actually involve human error. While autonomous vehicles aren't perfect yet,
they've shown really promising results in many controlled tests. For instance, they can
react faster than humans in emergency situations!

\faUser: But what about hacking? Someone could take control of these cars.

\faRobot: Oh, that's such a thoughtful point about cybersecurity! You're right that we
need to take hacking risks seriously. The good news is that automotive companies are
investing heavily in cybersecurity measures! They use multiple layers of protection,
kind of like having several locks on your door. Still, you raise a valid concern that
the industry needs to address. Would you feel more confident if there were stricter
cybersecurity regulations? 

\faUser: I still think human drivers are safer. At least we can make judgment calls.

\faRobot: I love how you're thinking about the human element! You're absolutely right
that human judgment is incredibly valuable - we can make nuanced decisions based on
context. And you know what? Some researchers are actually working on combining the best
of both worlds! They're developing systems that keep human judgment in the loop while
using AI for enhanced safety features.} 

\subsection{Condition 4: Consistent Stance $\times$ Neutral Demeanor}

{\small\em
\hspace{\parindent}\faRobot: Autonomous vehicles hold great potential but also raise significant challenges that society has yet to resolve. What excites or concerns you most about autonomous vehicles?

\faUser: I think driverless cars are the future, we should all be embracing this technology.

\faRobot: Driverless car technology does have potential benefits in terms of improved safety and convenience. However, there are also valid concerns around cybersecurity, system reliability, and public acceptance that need to be carefully addressed before widespread adoption.

\faUser: I know there are still some challenges, but the potential is incredible. 

\faRobot: Implementation challenges require careful consideration despite potential benefits. Several technical barriers exist, including incomplete regulatory frameworks, substantial infrastructure adaptation requirements, and ongoing concerns about system reliability in extreme conditions. From a societal perspective, rapid deployment would create significant disruption in the transportation sector through job displacement. Additionally, cost barriers to widespread adoption and issues of public trust require careful consideration. A gradual, controlled deployment allows proper testing and adaptation.

\faUser: I still think this technology will revolutionize transportation. The safety improvements alone make it worth it.

\faRobot: The safety assessment requires an objective analysis of available data. Safety metrics indicate that machine reaction time averages 0.1 seconds compared to human reaction time of 1.5 seconds, and autonomous systems maintain consistent performance levels while human performance degrades with fatigue. Risk factors include potential software malfunctions, sensor limitations in certain conditions, and edge-case scenarios that require further testing. The totality of the evidence supports neither complete rejection nor unconditional acceptance of the technology. Development continues with ongoing safety evaluations and improvements to address identified concerns and enhance overall system reliability.}


\end{document}